\documentclass[lettersize,journal]{IEEEtran}
\usepackage{amsmath,amsfonts}
\usepackage{amssymb}
\usepackage{algorithm}
\usepackage{algorithmicx}
\usepackage{array}
\usepackage[caption=false,font=footnotesize,labelfont=rm,textfont=rm]{subfig}
\usepackage{textcomp}
\usepackage{stfloats}
\usepackage{url}
\usepackage{verbatim}
\usepackage{graphicx}
\usepackage{cite}
\usepackage{ulem}
\usepackage{threeparttable}
\usepackage{newtxmath} 
\usepackage{enumitem}  

\usepackage[
pdfauthor={derajan},
pdftitle={How to do this},
pdfstartview=XYZ,
bookmarks=true,
colorlinks=true,
linkcolor=blue,
urlcolor=blue,
citecolor=blue,
pdftex,
bookmarks=true,
linktocpage=true,   
hyperindex=true
]{hyperref}

\begin{document}

\title{Drone Remote Identification Based on Zadoff-Chu Sequences and Time-Frequency Images}

\author{Jie Li,
	Jing Li,~\IEEEmembership{Member,~IEEE,}
	Lu Lv,~\IEEEmembership{Member,~IEEE,}
	Peixin Zhang,
	Fengkui Gong,~\IEEEmembership{Member,~IEEE,}
		\thanks{ \textit{(Corresponding author: Jing Li.)}}
		\thanks{Jie Li, Jing Li, Peixin Zhang, and Fengkui Gong are with the State Key Laboratory of Integrated Services Network, Xidian University, Xi'an, Shaanxi 710071, China (e-mail: lijie\_372@stu.xidian.edu.cn; jli@xidian.edu.cn; pxzhang@stu.xidian.edu.cn; fkgong@xidian.edu.cn).}
		\thanks{Lu Lv is with the School of Telecommunications Engineering, Xidian University, Xi'an 710071, China, and also with the Shaanxi Key Laboratory of Information Communication Network and Security, Xi'an University of Posts \& Telecommunications, Xi'an 710121, China (e-mail: lulv@xidian.edu.cn).}
}

\maketitle

\begin{abstract}
We propose an algorithm based on Zadoff-Chu (ZC) sequences and time-frequency images (TFI) to achieve drone remote identification (RID). Specifically, by analyzing the modulation parameters and frame structures of drone ratio-frequency (RF) signals in the DroneRFa dataset, we extract prior information about ZC sequences with surprising correlation properties and robustness. Cross-correlation is performed between locally generated ZC sequences and drone signals to derive ZC sequence-based features. Then, these ZF sequence features are fused with TFI features containing communication protocol information to achieve drone RID. To reduce computational costs, data reduction of the cross-correlation features is performed by analyzing the frame structures and modulation parameters, ensuring that the feature performance remained unaffected. Three feature fusion methods, namely probability-weighted addition, feature vector addition, and feature vector concatenation, are analyzed. Simulation results demonstrate that the proposed algorithm improves the average accuracy by at least 2.5\% compared to existing methods, which also indicate robust RID performance under burst interference and background noise. For RF sampling signals at varying flight distances, the proposed algorithm achieves a maximum accuracy of 99.11\%.
\end{abstract}


\section{Introduction}
\label{sec1}
\IEEEPARstart{U}{nmanned} aerial vehicles, commonly known as drones, have rapidly emerged as a transformative technology across various areas. From recreational applications and aerial photography to agriculture, logistics, and military operations, drones have demonstrated significant versatility and utility\textcolor{blue}{\cite{ref1}}. In 2022, the global drones fleet comprised 9.64 million units with a market size of 27.43 billion dollars, and this market is projected to grow from 31.7 billion dollars in 2023 to 91.23 billion dollars by 2030\textcolor{blue}{\cite{ref2}}. Unfortunately, the increasing proliferation of drones, particularly in densely populated and sensitive areas, has raised concerns about security, privacy, and regulatory compliance\textcolor{blue}{\cite{ref3}},\textcolor{blue}{\cite{ref4}},\textcolor{blue}{\cite{ref5}},\textcolor{blue}{\cite{ref6}},\textcolor{blue}{\cite{ref7}}. Some countries have developed registration, operation, and certification policies to mitigate potential risks associated with unauthorized or malicious drone activities\textcolor{blue}{\cite{ref8}}. However, existing works, such as micro-Doppler analysis of radar echoes\textcolor{blue}{\cite{ref9}},\textcolor{blue}{\cite{ref10}}, visual-based detection\textcolor{blue}{\cite{ref11}},\textcolor{blue}{\cite{ref12}}, acoustic-based detection\textcolor{blue}{\cite{ref13}},\textcolor{blue}{\cite{ref14}}, and time-domain feature transformations in radio-frequency (RF) signals\textcolor{blue}{\cite{ref15}}, have overlooked the potential benefits of leveraging communication protocols for drones remote identification (RID). On the other hand, methods utilizing modulation parameters\textcolor{blue}{\cite{ref16}} or frame structures\textcolor{blue}{\cite{ref17}} are not only vulnerable to burst interference but also fail to account for the inherent variability of signal parameters within the same frame. Thus, acquiring prior knowledge of modulation parameters by analyzing communication protocols, and selecting more effective and robust RF signal features, are imperative for achieving drones RID.

\subsection{Related Works}
\begin{enumerate}[leftmargin=0pt, itemindent=2pc, listparindent=\parindent]
	\item{ \textit{Non-Communication Protocol-Based}: Existing RID methods can be divided into four categories, that is, radar-based, visual-based, acoustic-based, and RF-based methods. Radar-based RID  adopts radar echoes to detect drones, but the small size and radar cross-section of miniature drones make them difficult to detect\textcolor{blue}{\cite{ref18}},\textcolor{blue}{\cite{ref19}}. Although visual-based RID can leverage cameras to capture the images and videos of drones, it is challenging to guarantee the covering space and effectiveness for limited cameras and discontinuous data\textcolor{blue}{\cite{ref20}},\textcolor{blue}{\cite{ref21}}. Acoustic-based RID adopts acoustic features from propeller and motor to identify and track drones, but has poor robustness under low signal-to noise ratio (SNR) situations\textcolor{blue}{\cite{ref22}},\textcolor{blue}{\cite{ref23}},\textcolor{blue}{\cite{ref24}},\textcolor{blue}{\cite{ref25}}. 
		
	The study of RF signals between drones and pilots is a promising solution for drones RID, that is because RF signals are not constrained by radar cross-section, low-light scenarios, and non-line-of-sight situations. The work in\textcolor{blue}{\cite{ref26}} proposed a multi-stage algorithm, where SNR threshold for detecting RF signals, modulation parameters for identifying drones signals, and 15 statistical RF features for classifying 17 drones. However, the bandwidths for drones in the considered database are all less than 10 MHz, which makes this multi-stage algorithm unsuitable for identifying drones with bandwidth exceeding 10 MHz from wireless fidelity (Wi-Fi) interference. A drones RID algorithm based on in-phase/quadrature (I/Q) sequences was proposed to identify drones and detecting drones' operation modes, which adopted blockchain and edge computing to guarantee secure and real-time detection\textcolor{blue}{\cite{ref27}}. By using array antennas to collect RF signals, the signal frequency spectrum, wavelet energy entropy, and power spectral entropy were extracted to detection drones. Additionally, the angle of azimuth and elevation were computed from channel state information (CSI) to position drones\textcolor{blue}{\cite{ref28}}. Utilizing time-frequency images (TFI) can also achieve drone RID\textcolor{blue}{\cite{ref29}},\textcolor{blue}{\cite{ref30}},\textcolor{blue}{\cite{ref31}}, signal detection\textcolor{blue}{\cite{ref32}}, and denoising\textcolor{blue}{\cite{ref33}},\textcolor{blue}{\cite{ref34}} modules can also be introduced integrated into TFI-based drones RID frameworks.
	}

	\item{ \textit{Communication Protocol-Based}: Since drones' uplink frequency hopping spread spectrum (FHSS)\textcolor{blue}{\cite{ref35}} and downlink orthogonal frequency division multiplexing (OFDM)\textcolor{blue}{\cite{ref36}} signals can reveal modulation parameters and frame structures, several works tried to directly extract the communication protocol information contained in I/Q sequences. Different from the 8 features in\textcolor{blue}{\cite{ref37}}, 12 features were adopted from drones and non-drones signals to describe the attributes of packet size and packet inter-arrival time, which can solve the RID problem of 8 drone types and 8 operation patterns separately\textcolor{blue}{\cite{ref17}}. Considering that preambles will be added to the begin of downlink signal frames for synchronizing and estimating CSI, I/Q and cross correlation sequences were utilized to identify 4 drone types under static and hovering scenarios\textcolor{blue}{\cite{ref15}}.\textcolor{blue}{\cite{ref16}} estimated the modulation parameters e.g., the length of cyclic prefix (CP) and OFDM data symbols, to compute the normalized cyclic prefix correlation spectrum (NCPCS) and identify drones. However, it should be mentioned that in the process of calculating NCPCS, not only the CP of data symbols but also the CP of preambles can be utilized to achieve more accurate RID. Besides, similar to the frame structures in long term evolution (LTE), the CP length of different OFDM symbols even in the same frame is not completely fixed, i.e., normal CP and extended CP\textcolor{blue}{\cite{ref38}}. Thus, it is not always possible to estimate CP length via calculating the cyclic auto-correlation function directly.}
	
\end{enumerate}

\subsection{Motivations and Contributions}
Although prior works have explored drones RID via various RF-based algorithms, there still exist several challenges in communication protocol-based RID, as follows:

\begin{itemize}
	\item{The exploration of downlink signal frame structures is insufficient. Few works have noticed the variability of CP length even in the same frame, which will result in an inaccurate frame structure reconstruction without distinguishing OFDM symbols. Specifically, inaccurate estimation of CP length will lead to errors in estimating the starting and ending positions of symbols, which will ulteriorly affect the effectiveness of extracted feature quantities for drones RID. Besides, to our knowledge, there is no prior work in the literature that analyzes and utilizes the preamble sequence to achieve drones RID. The preambles, usually Zadoff-Chu (ZC) sequences, have surprising auto-correlation and cross-correlation characteristics, which ensure the RID accuracy under low SNR scenarios and in the presence of interference.
	}
	\item{Utilizing uplink signal modulation parameters to achieve drones RID is susceptible to burst interference. Since the frequencies of uplink signals contain 845 MHz, 1430 MHz, 2.4 GHz, and 5.8 GHz, there exists bluetooth, Wi-Fi, and other burst interference when collecting RF samples from drones\textcolor{blue}{\cite{ref34}}. On the one hand, when there is limited samples, the periodicity of FHSS signals cannot be reflected. On the other hand, when there is a large amount of samples, burst interference will block some FHSS signals and weaken the continuity.}
\end{itemize}

Motivated by the above considerations, we investigate the drones RID problem on an outdoor practical sampling dataset named DroneRFa\textcolor{blue}{\cite{ref39}}. Specifically, the modulation parameters and frame structures of drones RF signals are analyzed to obtain prior information of ZC sequences. Since TFI can reveal modulation parameters and frame structures for OFDM and FHSS signals to a certain extent, we compute and fuse the TFI and cross-correlation results of ZC sequences to achieve drones RID. 
The contributions of this paper are summarized as follows.

\begin{itemize}
	\item{We first analyze and utilize ZC sequences to identify drones. By analyzing the modulation parameters and frame structures, we conduct an investigation into the potential ZC sequences utilized by 8 types of drones. Different drones employ distinct ZC sequences, and the surprising auto-correlation and cross-correlation properties of ZC sequences make them a natural choice for RID. Based on the prior knowledge, we extracted their features by computing the cross-correlation sequences between I/Q samples and the locally generated identical sequences. To reduce computational costs, a data reduction of the sequences is performed to obtain cross-correlation features by analyzing the frame structures and parameters, without degrading the feature performance.
		}
	\item{The impact of different feature fusion methods between TFI and cross-correlation features are explored. TFI can partially reveal the modulation parameters and frame structures of both uplink and downlink drones signals, and provide additional information that ZC sequences cannot capture. Meanwhile, cross-correlation features exhibit robust RID accuracy under low SNR conditions and in the presence of burst interference, compensating for the drawbacks of TFI images being susceptible to noise and interference. Consequently, we send both TFI and cross-correlation features into the neural network and investigate the impact of three fusion methods on RID accuracy, i.e., probability-weighted addition (PWA), feature vector addition (FVA), and feature vector concatenation (FVC).}
	\item{Numerical results demonstrate that the proposed algorithm effectively achieves drone RID on DroneRFa. Specifically: 1) Compared with the baseline algorithms, the proposed algorithm based on ZC sequences and TFI improves the average accuracy by at least 2.5\%, validating the effectiveness of the selected features for drone RID. 2) Among the evaluated feature fusion methods, FVC outperforms PWA and FVA in both computational cost and accuracy. 3) For drones operating at varying flight distances, the proposed algorithm achieves an accuracy exceeding 80\% when SNR = -15 dB.}
\end{itemize}

\subsection{Outline and Notations}
The signal analysis and feature extraction are presented in Section \ref{sec2}. The details and explanations of drones RID algorithm are discussed in Section \ref{sec3}. Simulation evaluation is presented in Section \ref{sec4}. And this paper is summarized in Section \ref{sec5}.

\section{Signal Analysis and Feature Extraction}
\label{sec2}
\subsection{Signal Analysis}

\begin{figure}[!t]
	\centering
	\includegraphics[width=2.5in]{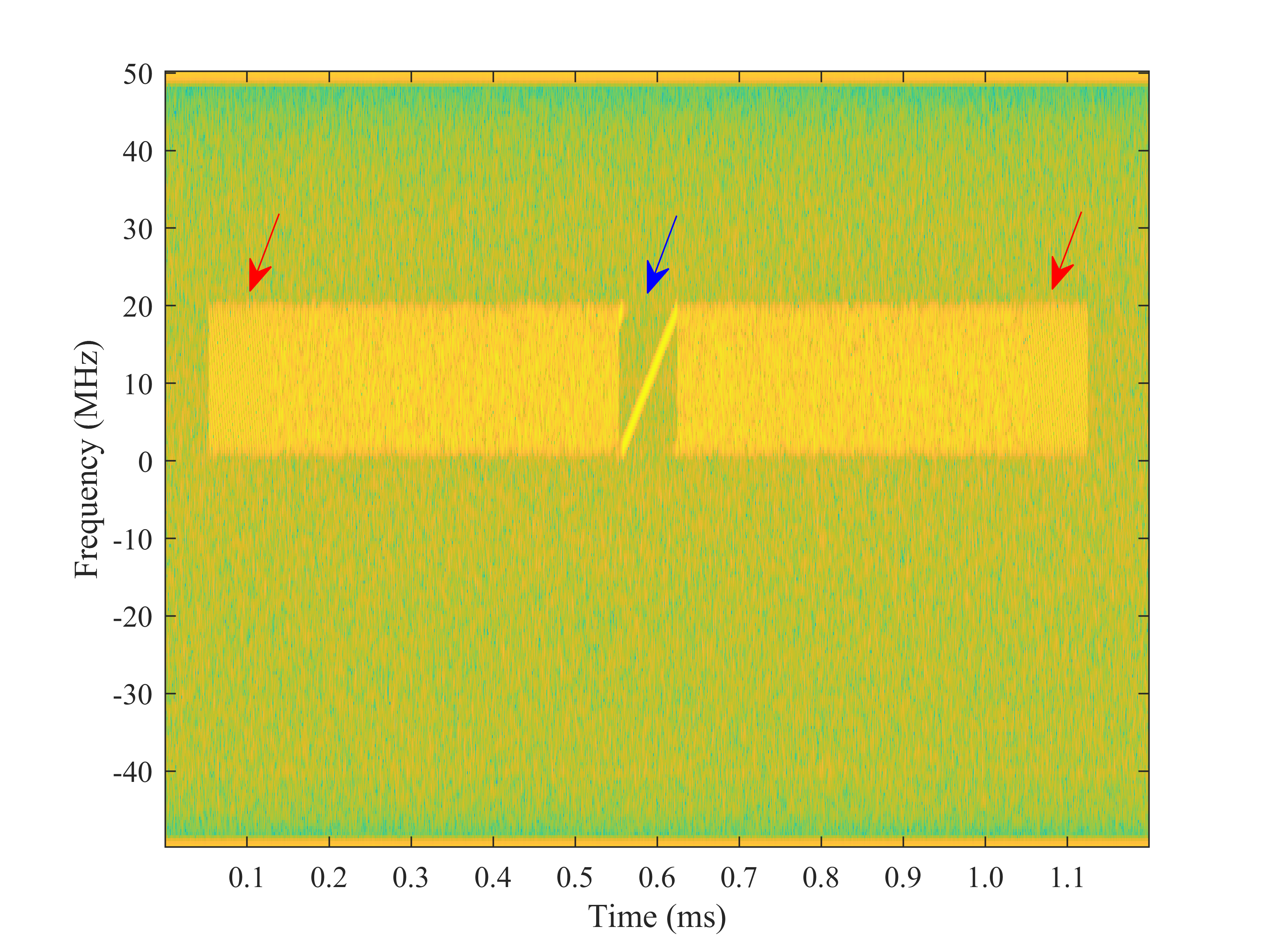}
	\caption{TFI of one signal frame of T0010D00.}
	\label{Fig_1}
\end{figure}

The drones RF signals dataset, named DroneRFa, is sampled at 100 MHz across three open frequency bands, i.e., 915 MHz, 2.4 GHz, and 5.8 GHz. Based on the analysis of frame structures, 8 types of drones adopt ZC sequences are selected, and the interpretation of the binary labels is provided in Table \ref{tab:table1}.
Considering the sampling rate \(f_s=100\) MHz, storing samples over longer sampling duration poses challenges. Therefore, we select only 100 ms of I/Q sequences as the raw samples. Taking the DJI Mini 3 Pro as an example, a typical frame consists of 15 OFDM symbols with 2048 sub-carriers, spanning approximately 1 ms. Hence, 100 ms of \(L=10^7\) raw samples contain sufficient protocol information to achieve drones RID. TFI of one signal frame of T0010D00 is simulated as shown in Fig. \ref{Fig_1}, the OFDM symbols pointed by the red and blue arrows correspond to different ZC sequences, respectively.
\begin{table}[!t]   
	\caption{Explanation of the Dataset's Labels}  
	\label{tab:table1} 
	\centering
	\begin{threeparttable}
		\begin{tabular}{|c|c|c|}   
			\hline   \textbf{Label1} & \textbf{Labe2} & \textbf{Drone Types} \\   
			\hline   T0000 & D00 & Background Noise and Interference \\ 
			\hline   T0001 & D00 D01 D10 & DJI Air 2S \\  
			\hline   T0010 & D00 & DJI Mini 3 Pro \\
			\hline   T0011 & D00 & DJI Mavic Pro \\
			\hline   T0100 & D00 D01 D10 & DJI Mini 2 \\ 
			\hline   T0101 & D00 & DJI Mavic 3 \\ 
			\hline   T0110 & D00 & DJI MATRICE 300 \\ 
			\hline   T0111 & D00 & DJI Phantom 4 Pro RTK\\
			\hline   T1000 & D00 & DJI MATRICE 30T \\
			\hline   
		\end{tabular}
		\begin{tablenotes}
			\footnotesize
			\item[1] D00, D01 D10 denote the flight distance range of 20\(\sim\)40, 40\(\sim\)80, 80\(\sim\)150 m, respectively.
		\end{tablenotes}            
	\end{threeparttable}  
\end{table}

Let \(x(l)\) with \(l \in [0, 1, \ldots, L-1]\) denote the I/Q complex samples, the auto-correlation sequences for \(N_{u\!p}\) samples can be obtained by
\begin{equation}
	\label{Eq1}
	\gamma ( m )\text{=}\left| \sum\limits_{k=1}^{{{N}_{u\!p}}}{x( k ){{x}^{*}}( k+m )} \right|,m =0,1,\ldots ,L-{{N}_{u\!p}-1}.
\end{equation}

Since the modulation parameters and frame structures of drone signals are fixed after manufacturing, certain parameters can be estimated and utilized as prior knowledge for drone RID. To calculate the signal bandwidth \(B\), \(x(l)\) will be divide the sample into \(K\) overlapping data segments, which can be expressed as
\begin{equation}
	\label{Eq2}
	{{x}_{i}}(l)=x(iD+l)w(n), 0 \le i \le K-1,
\end{equation}
where \(w(n)\) is a window with a duration of \(L_w\), used to control spectral leakage. \(D\) is the offset length, where  \(D \le L_w\). The periodogram of the \(i\text{-th}\) segment can be given by
\begin{equation}
	\label{Eq3}
	{{R}_{x,i}}({{e}^{j\omega }})\triangleq \frac{1}{{{L}_{w}}}{{\left| {{X}_{i}}({{e}^{j\omega }}) \right|}^{2}}=\frac{1}{{{L}_{w}}}{{\left| \sum\limits_{l=0}^{L-1}{{{x}_{i}}(l){{e}^{-j\omega l}}} \right|}^{2}}.
\end{equation}

Assuming \(D = L_w\), i.e., the data segments do not overlap, the spectral estimate can be obtained by averaging \(K\) periodograms, represented as
\begin{equation}
	\label{Eq4}
	{{R}_{x}}({{e}^{j\omega }})\triangleq \frac{1}{K}\sum\limits_{i=0}^{K-1}{{{R}_{x,i}}({{e}^{j\omega }})}=\frac{1}{K{{L}_{w}}}\sum\limits_{i=0}^{K-1}{{{\left| {{X}_{i}}({{e}^{j\omega }}) \right|}^{2}}}.
\end{equation}

A more commonly used method is to set \(D = \frac{L_w}{2}\), in which case \({{R}_{x}}({{e}^{j\omega }})\) can be given by
\begin{equation}
	\label{Eq5}
	{{R}_{x}}({{e}^{j\omega }})=\frac{1}{K{{L}_{w}}}{{\left| \sum\limits_{l=0}^{L-1}{{{x}_{i}}(l){{e}^{-j\omega l}}} \right|}^{2}}.
\end{equation}

\begin{figure}[!t]
	\centering
	\includegraphics[width=2.5in]{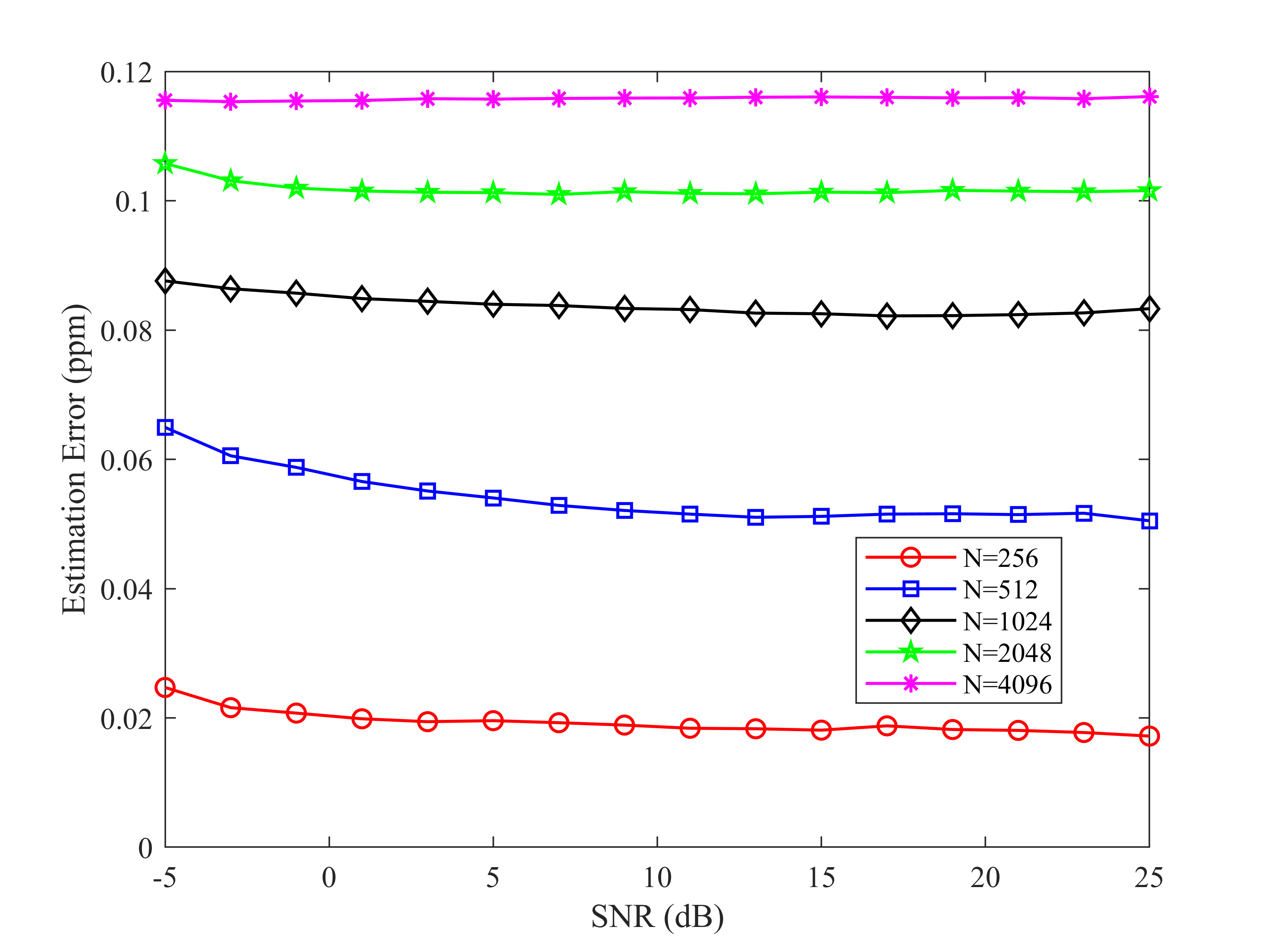}
	\caption{The estimation error of bandwidth versus \(\tiny{N}\) for integer sampling.}
	\label{Fig_2}
\end{figure}

\begin{figure}[!t]
	\centering
	\includegraphics[width=2.5in]{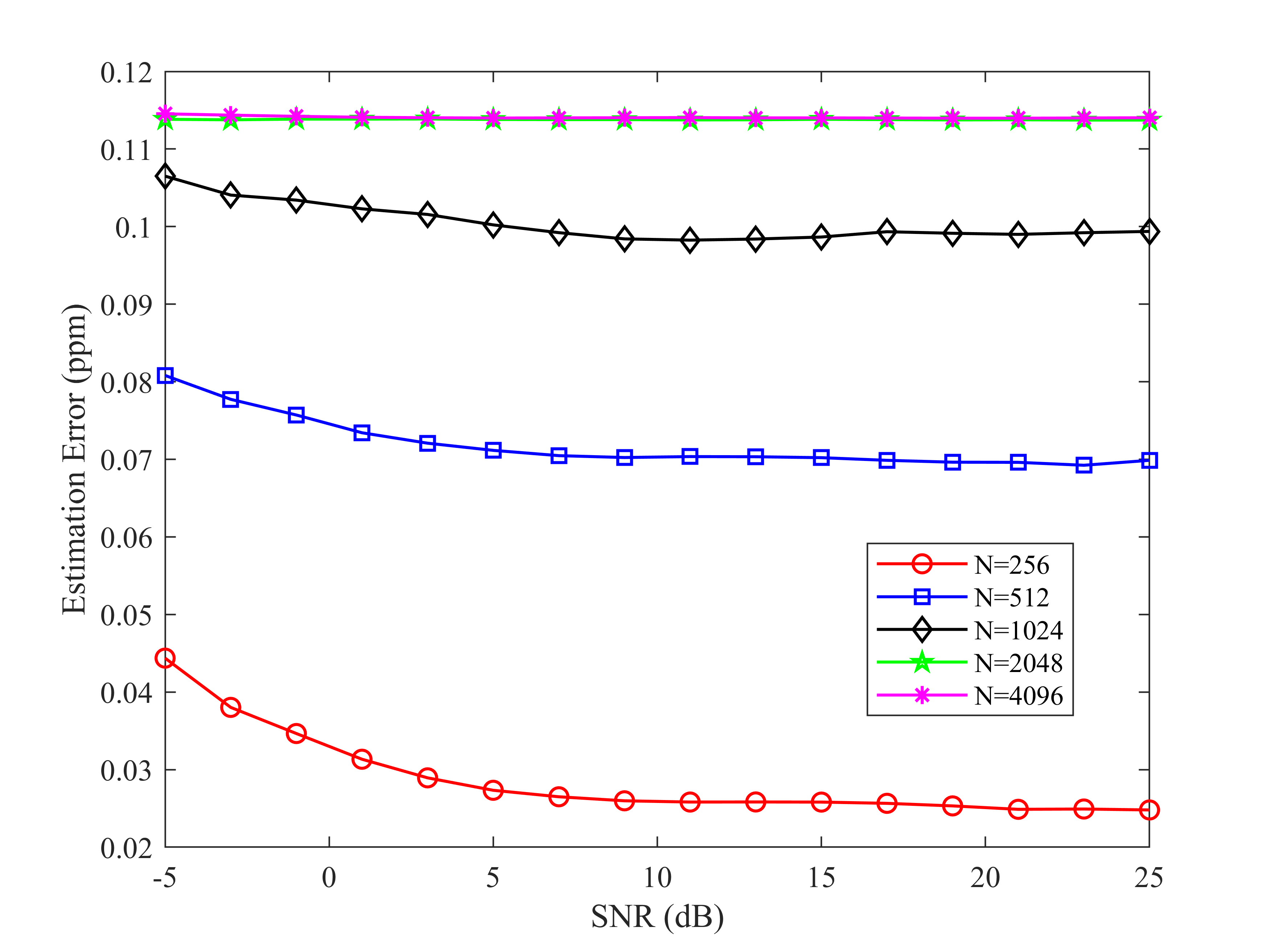}
	\caption{The estimation error of bandwidth versus \(\tiny{N}\) for non-integer sampling.}
	\label{Fig_3}
\end{figure}

By overlapping 50\% of the data segments, \(K\) becomes twice as large, which in turn reduces the variance of the periodogram by 50\%. Further overlap does not lead to additional reduction in variance, as the independence between the data segments decreases\textcolor{blue}{\cite{ref40}}. For \({{R}_{x}}({{e}^{j\omega }})\), the estimated bandwidth \(\hat{B}_v\) is typically calculated at the position where the attenuation is 3 dB. OFDM signals with a bandwidth of 20 MHz over multipath channels are generated, and the bandwidth estimation is performed with 4-times integer sampling and \(\frac{37}{9}\)-times non-integer sampling. The simulation results are shown in Fig. \ref{Fig_2} and Fig. \ref{Fig_3}, it can be observed that non-integer sampling will degrade the performance of bandwidth estimation. For commonly used OFDM signals with a sum number of subcarriers of 1024 and 2048, the bandwidth estimation error does not exceed 120 KHz at most. Considering that OFDM signals are typically implemented with a fixed number of \(N\), the bandwidth can essentially be estimated with 100\% accuracy after rounding to the nearest MHz.


Besides, it is important to note that, because virtual subcarriers are typically inserted during OFDM systems, the estimated bandwidth \(\hat{B}_v\) needs further analysis and calculation to obtain the actual bandwidth \(B\). Specifically, taking the T0010D00 signal as an example, the bandwidth estimation operation yields \(\hat{B}_v = 18\) MHz, which aligns with the direct observation results on the TFI shown in Fig. \ref{Fig_1}. However, upon closer analysis, it is determined that the bandwidth \(\hat{B}_v\) should be 18.015 MHz, corresponding to 1201 subcarriers, which includes 1200 data subcarriers and 1 DC subcarrier. The remaining 847 virtual subcarriers are distributed symmetrically on both sides of the data subcarriers, with 424 on one side and 423 on the other. Therefore, \(N\) should be 2048, corresponding to a bandwidth of \(B = 30.72\) MHz.

\begin{figure}[!t]
	\centering
	\includegraphics[width=2.5in]{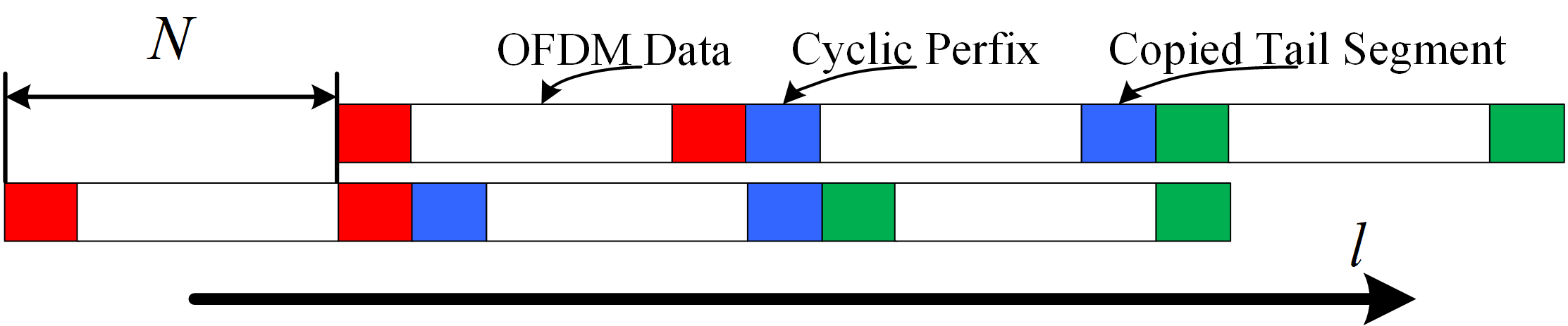}
	\caption{Graphical explanation for auto-correlation of OFDM signals.}
	\label{Fig_4}
\end{figure}

\begin{figure}[!t]
	\centering
	\includegraphics[width=2.5in]{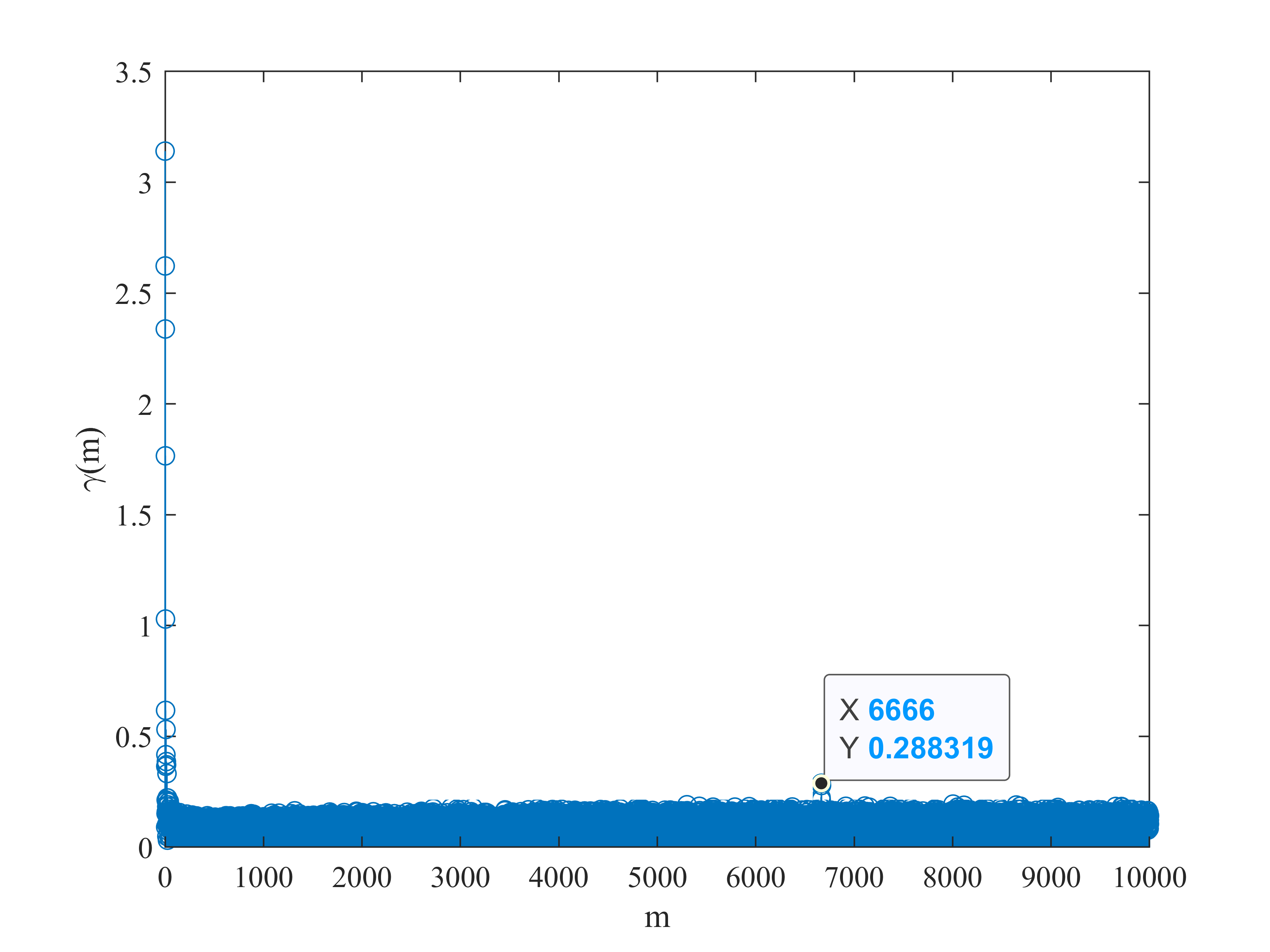}
	\caption{Simulated value of \(\tiny{\gamma(m)}\) for T0010D00.}
	\label{Fig_5}
\end{figure}

The bandwidth estimation can also be cross-validated with the estimation of the sum number of subcarriers. For OFDM signals, \(\gamma(m)\) in \textcolor{blue}{(\ref{Eq1})} will exhibit a peak at \(\frac{N{{f}_{s}}}{B}\) as shown in Fig. \ref{Fig_4} and Fig. \ref{Fig_5}. Let \(m^*\) denote the index of the maximum value of \(\gamma(m)\), there should have
\begin{equation}
	\label{Eq6}
	\frac{{{m}^{*}}}{{{f}_{s}}}=\frac{N}{B}.
\end{equation}

\begin{figure}[!t]
	\centering
	\includegraphics[width=2.5in]{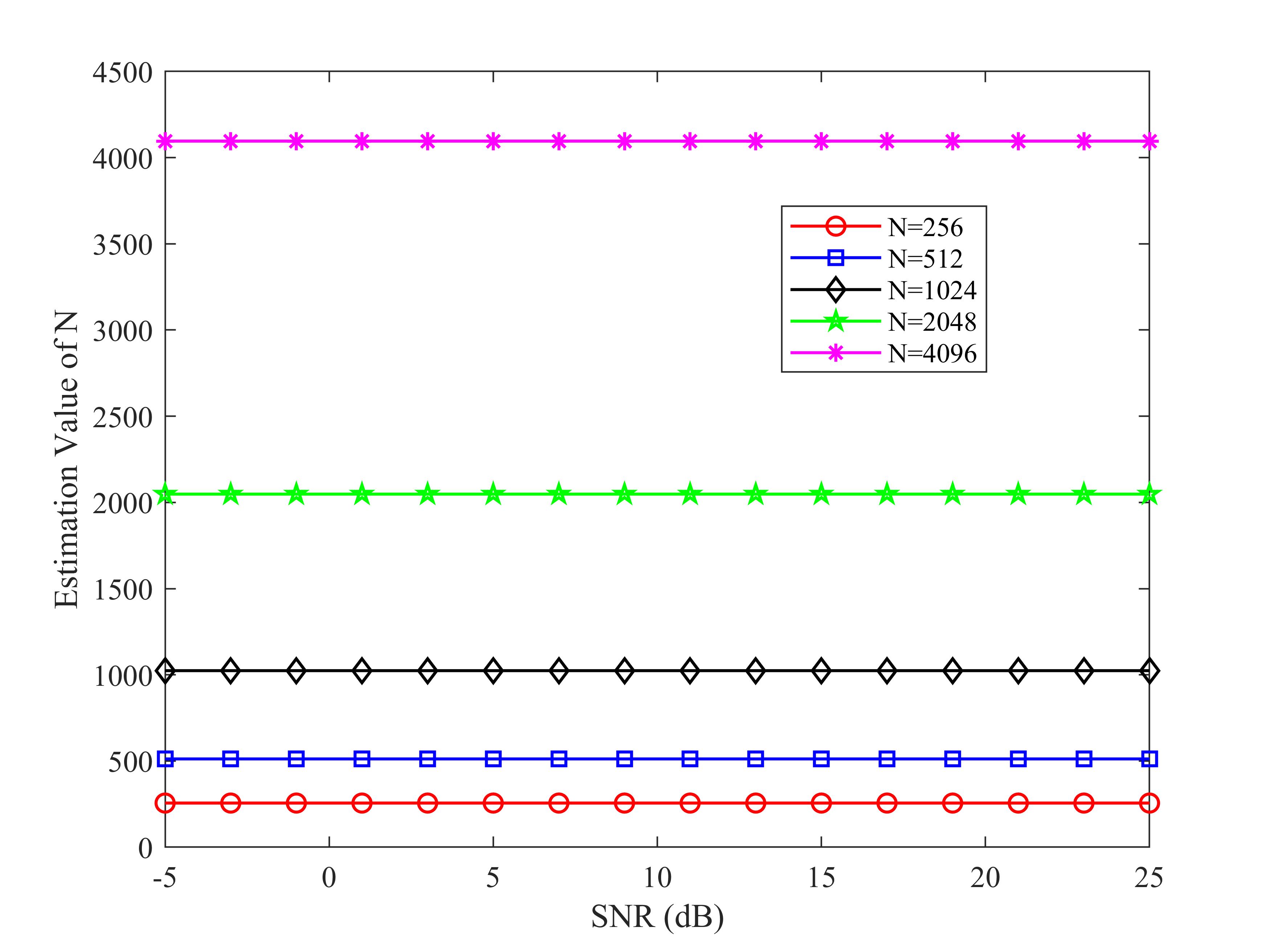}
	\caption{The estimation value of \(\tiny{N}\) versus SNR for integer sampling.}
	\label{Fig_6}
\end{figure}

\begin{figure}[!t]
	\centering
	\includegraphics[width=2.5in]{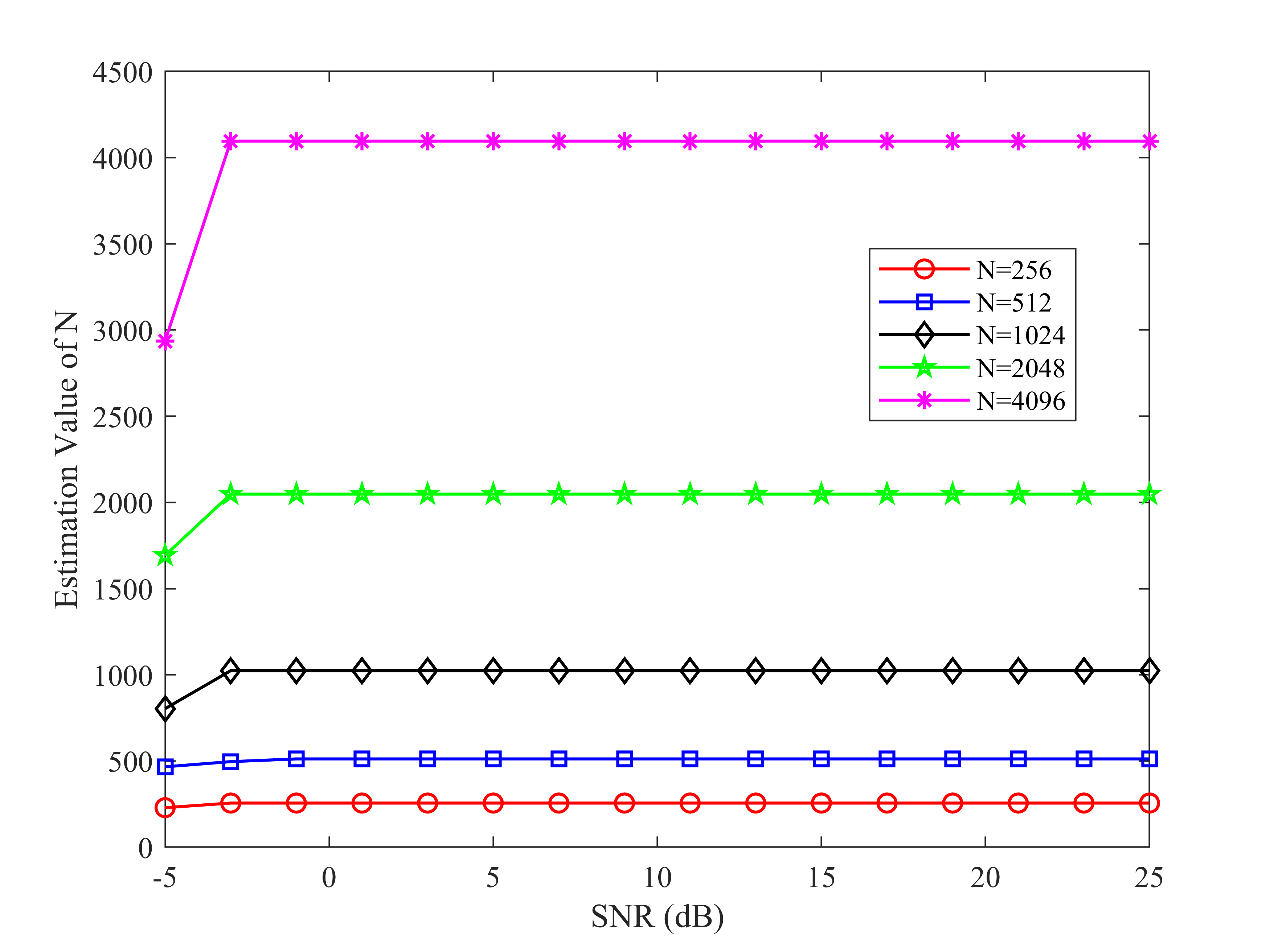}
	\caption{The estimation value of \(\tiny{N}\) versus SNR for non-integer sampling.}
	\label{Fig_7}
\end{figure}

Besides, \textcolor{blue}{(\ref{Eq1})} is performed for the same OFDM signals in bandwidth estimation, and the estimation results are shown in Fig. \ref{Fig_6} and Fig. \ref{Fig_7}. To facilitate the implementation of OFDM modulation with fast Fourier transform, \(N\) is typically a power of 2, such as 1024 or 2048. Simulation results indicate that although non-integer sampling may lead to less accurate estimations of \(N\) at low SNR, it remains entirely accurate and feasible when analyzing OFDM signal parameters at high SNR.


Based on the above analysis, signal bandwidth \(B\) and the number of sum sub-carriers \(N\) can be accurately obtained through Welch and auto-correlation method.

\subsection{Feature Extraction}
\begin{enumerate}[leftmargin=0pt, itemindent=2pc, listparindent=\parindent]
	\item{ \textit{TFI Features}: Short-time Fourier transform is adopted to calculate TFI of RF signals, the value of \(t\text{-th}\) time index and \(f\text{-th}\) frequency index is given by
	\begin{equation}
		\label{Eq7}
		X( t,f )\text{=}\left| \sum\limits_{\tau =-\infty }^{+\infty }{x( \tau ){{h}^{*}}( \tau -t ){{e}^{-j2\pi f\tau }}} \right|,
	\end{equation}
	where \(h(\tau)\) denotes the window function, the window length \(W\) is equal to the fast Fourier transform size. For \(L=10^7\) raw samples, the approximate computational complexity is \(\mathcal{O}( L\log W )\), which is acceptable. Fig. \ref{Fig_8} shows the TFI for \(10^7\) samples of T0100D00, it can be seen that TFI contains the communication protocol information both of OFDM and FHSS.

\begin{figure}[!t]
	\centering
	\includegraphics[width=2.5in]{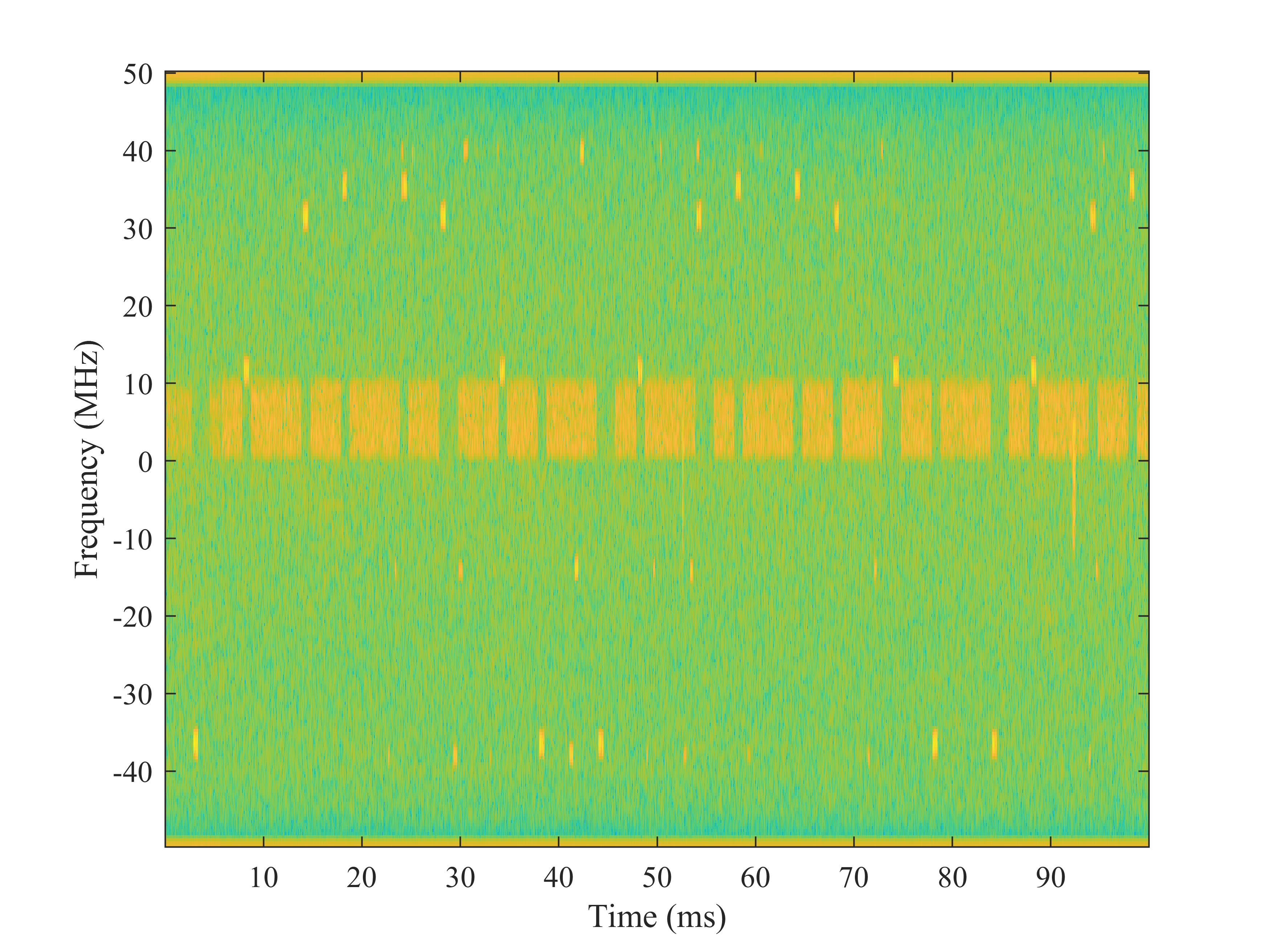}
	\caption{TFI of T0100D00 with \(\tiny{10^7}\) samples at SNR = 15 dB.}
	\label{Fig_8}
\end{figure}
}
	
	\item{ \textit{Cross-Correlation Features for ZC Sequences}: ZC sequences are widely employed in synchronization and channel estimation, which have found extensive applications in various wireless communication systems, including LTE , worldwide interoperability for microwave access, and global navigation satellite system. ZC sequences can be generated by
	\begin{equation}
		\label{Eq8}
		{{z}_{u}}( i )={{e}^{-j\frac{\pi ui(i+1)}{{{L}_{z\!c}}}}},i=0,1,\ldots ,{{L}_{z\!c}}-1,
	\end{equation}
	where \(u\) and \(L_{z\!c}\) denote the physical root index and sequence length, respectively, and there is regulation that \(u\le {{L}_{z\!c}}-1\). Notably, it is typically the case that \(L_{z\!c} = N - N_v + 1\) in OFDM systems. For ZC sequences with the same \(u\), the cross-correlation operation is performed between the sequence itself and the sequence shifted by \(N_s\) points, resulting in a peak amplitude \(L_{z\!c}\) only at the \(N_s\text{-th}\) point, with other points approximately equal to zero. For ZC sequences with different \(u\) but the same \(L_{z\!c}\), if \(L_{z\!c}\) and \(\left| u_1-u_2 \right|\) have the greatest common divisor \(N_d\), there are peak amplitudes only at 	\(\left\{ i\left| i\le {{L}_{z\!c}},i\in {{\mathbb{Z}}^{+}} \right. \right\}\) with \(\sqrt{N_dL_{z\!c}}\), and other points are approximately equal to zero.
	
	Thus, it is straightforward to determine the values of \(u\) and \(L_{z\!c}\) employed by different types of drones, thereby enabling the local generation of identical ZC sequences. \textcolor{blue}{(\ref{Eq1})} can be rewritten as
	\begin{equation}
		\label{Eq9}
		\gamma ( m )\text{=}\left| \sum\limits_{k=1}^{{{N}_{u\!p}}}{y( k ){{x}^{*}}( k+m )} \right|,m=0,1,\ldots ,L-{{N}_{u\!p}},
	\end{equation}
	where \(y( k )\) denotes the sequence obtained by resampling the reconstructed OFDM symbol with \(z_u\). 
	
	However, the computational complexity for cross-correlation operation is \(\mathcal{O}( LN_{u\!p} )\), which fails to meet the real-time requirement for drone RID. Thus, data reduction is adopted to shorten the cross-correlation results, thereby reducing the cost of neural network. Specifically, through the analysis of frame structures, it is evident that ZC sequences are consistently present in 7 OFDM symbols of most drone signals. To mitigate the risk of missing ZC sequences, we randomly selected \(U=20\) non-overlapping segments, each of length \(V=5\times10^4\), from \(L=10^7\) samples to construct a new set of raw samples with \(L_{re}=UV=10^6\). Cross-correlation results of one signal frame of T0010D00 is simulated as shown in Fig. \ref{Fig_9}, and ZC1 \(\sim\) ZC8 represent the cross-correlation results of the selected ZC sequences employed by different types of drones. It can be observed that the local sequence represented by ZC7 exhibits a distinct cross-correlation peak with the signal from T0010D00, demonstrating the feasibility of utilizing ZC sequences for drone RID.
	\begin{figure}[!t]
		\centering
		\includegraphics[width=2.5in]{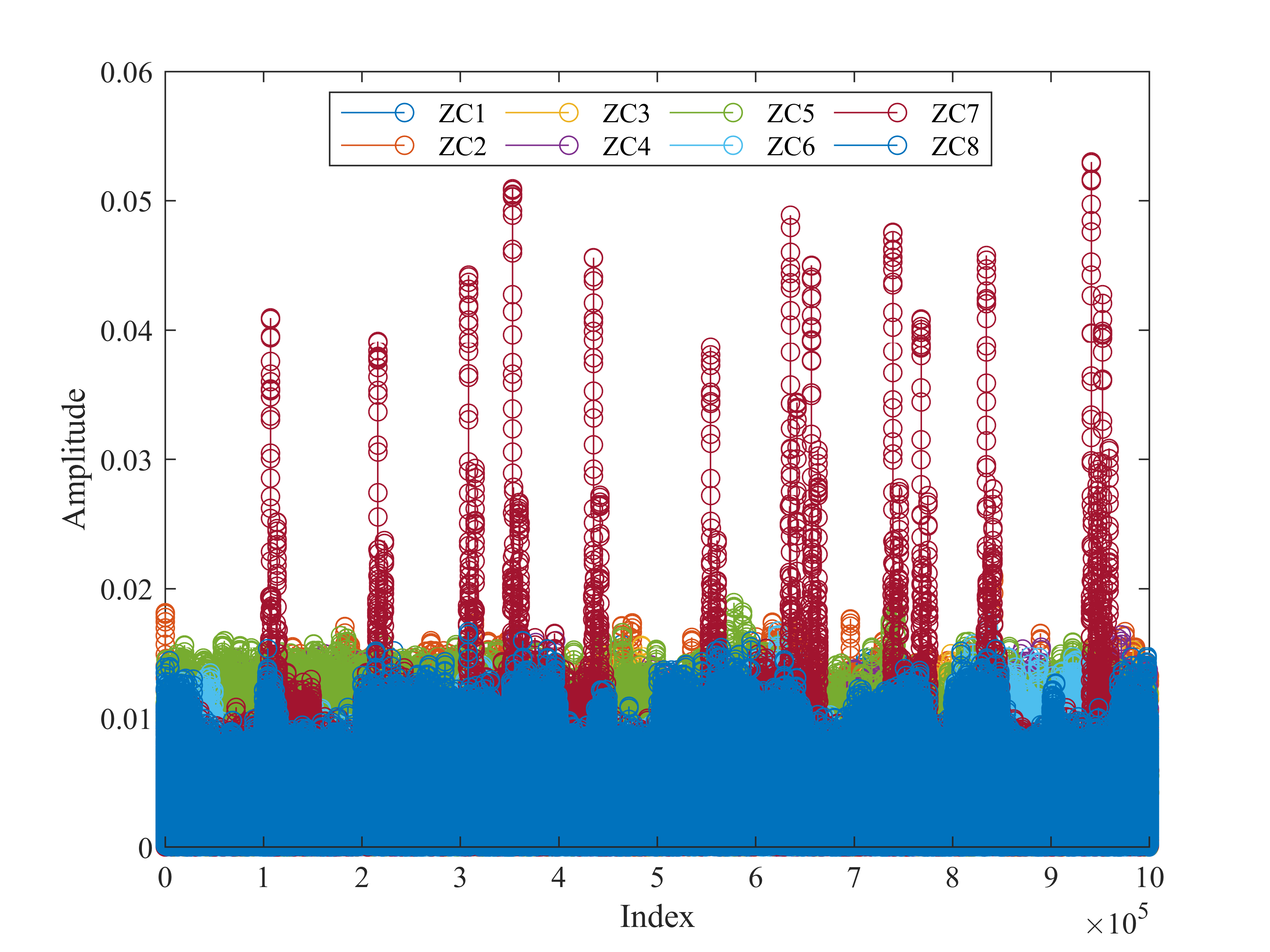}
		\caption{\(\tiny{\gamma_{re}}\) for different ZC sequences of T0010D00.}
		\label{Fig_9}
	\end{figure}
	\begin{figure}[!t]
		\centering
		\includegraphics[width=2.5in]{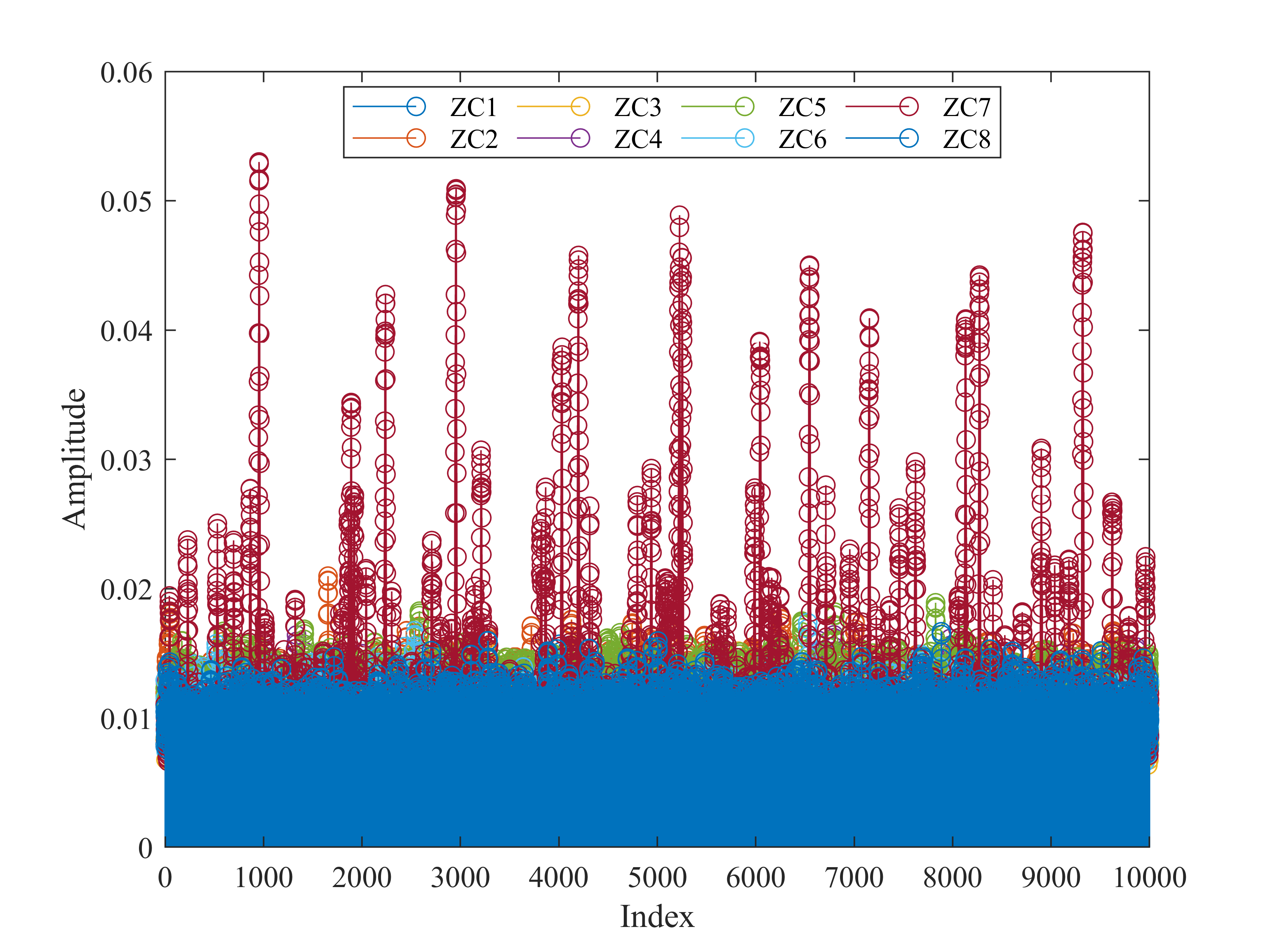}
		\caption{\(\tiny{\mathbf{R}_i}\) for different ZC sequences of T0010D00.}
		\label{Fig_10}
	\end{figure}
	
	Then \(\gamma_{r\!e}(m)\) in \textcolor{blue}{(\ref{Eq1})} will be reshaped to \(\mathbf{r}\in {{\mathbb{R}}^{U_1\times V_1}}\) with \(U_1=5U\) and \(V_1=\frac{V}{5}\), which can be give by
	\begin{equation}
		\label{Eq10}
		\mathbf{r}=\left[ \begin{matrix}
			{{\gamma}_{r\!e}}\!(1) & {{\gamma}_{r\!e}}\!( 2 ) & \cdots  & {{\gamma}_{r\!e}}\!( \frac{V}{5} )  \\
			{{\gamma}_{r\!e}}\!( \frac{V}{5}\!+\!1 ) & {{\gamma}_{r\!e}}\!( \frac{V}{5}\!+\!2 ) & \cdots  & {{\gamma}_{r\!e}}\!( \frac{2V}{5} )  \\
			\vdots  & \vdots  & \ddots  & \vdots   \\
			{{\gamma}_{r\!e}}\!( \frac{5U-1)V}{5}\!+\!1 ) & {{\gamma}_{r\!e}}\!(\frac{(5U-1)V}{5}\!+\!2) & \cdots  & {{\gamma}_{r\!e}}\!(UV)  \\
		\end{matrix} \right].
	\end{equation}
	
	The cross-correlation feature for \(i\text{-th}\) type of ZC sequence can be obtained by
	\begin{equation}
		\label{Eq11}
		\mathbf{R}_i=\left[ \begin{matrix}
			\max {{\mathbf{r}}_{1}} & \max {{\mathbf{r}}_{2}} & \cdots  & \max {{\mathbf{r}}_{V_1}}  \\
		\end{matrix} \right],
	\end{equation}
	where \({{\mathbf{r}}_{j}}\) denotes the \(j\text{-th}\) column of \(\mathbf{r}\). The results of data reduction are simulated as shown in Fig. \ref{Fig_10}. Intuitively, this operation seems to amplify the impact of certain interference on the cross-correlation results of ZC sequences. However, it merely highlights the details of the original results through coordinate scaling. Both \(\gamma_{r\!e}\) and \(\mathbf{R}_i\), when utilized as input sequence features for neural network, inherently include the interference. Furthermore, data reduction compresses the feature size by a factor of 10, preserving the prominent distinguishing peaks while significantly decreasing the number of parameters required for network training.
	
	As highlighted in the previous analysis, each of the 8 different drone types in the dataset is associated with a specific ZC sequence. Consequently, the final size and overall computational complexity for cross-correlation feature is \(8\times V_1\) and \(\mathcal{O}( 8UV( {{N}_{u\!p}}+1 ) )\), respectively.
	}
\end{enumerate}
 
\section{Drones RID Algorithm}
\label{sec3}

\begin{figure}[!t]
	\centering
	\includegraphics[width=2.5in]{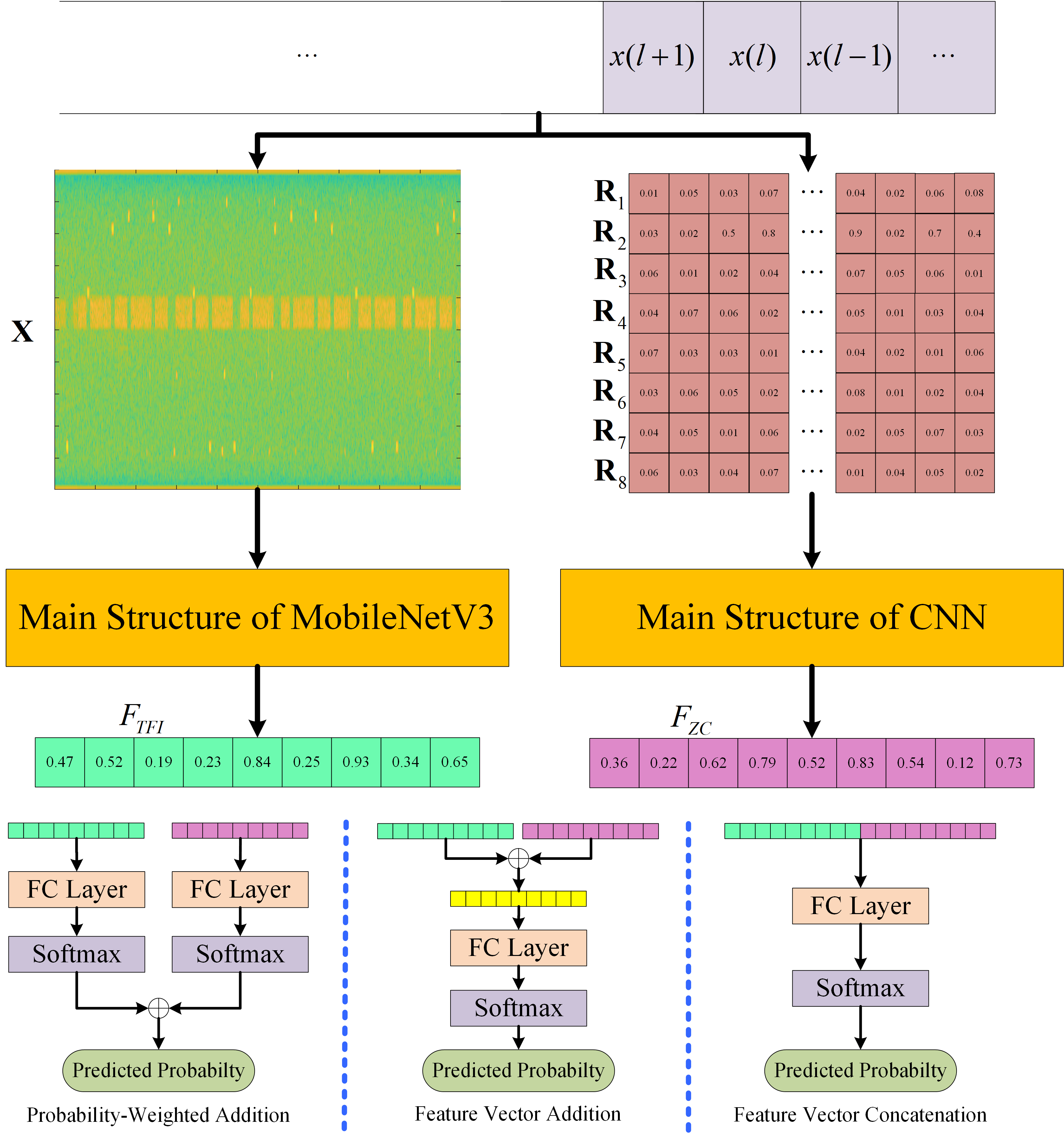}
	\caption{A sketch of the proposed drone RID algorithm.}
	\label{Fig_11}
\end{figure}

To highlight the impact of cross-correlation features on RID accuracy, we utilize the classic classification model, image-based MobileNetV3 and sequence-based CNN\textcolor{blue}{\cite{ref16}}, as the neural network architectures. Fig. \ref{Fig_11} illustrates the architecture and three feature fusion methods for the proposed drone RID algorithm.

Define \({{\mathbf{X}}_{t,f}}=X(t,f)\), the feature vectors output by MobileNetV3 and CNN can be expressed as
\begin{equation}
	\label{Eq12}
	F_{T\!F\!I} = {{f}_{T\!F\!I}}(\mathbf{X},{{\varpi }_{T\!F\!I}}),
\end{equation}
\begin{equation}
	\label{Eq13}
	F_{Z\!C} = {{f}_{Z\!C}}([{\mathbf{R}}_{1},{\mathbf{R}}_{2},{\mathbf{R}}_{3},{\mathbf{R}}_{4},{\mathbf{R}}_{5},{\mathbf{R}}_{6},{\mathbf{R}}_{7},{\mathbf{R}}_{8}],{{\varpi }_{Z\!C}}),
\end{equation}
where \(f_{T\!F\!I}\) and \(f_{Z\!C}\) denotes the the operations performed by MobileNetV3 and CNN, respectively, where \(\varpi_{T\!F\!I}\) and \(\varpi_{Z\!C}\) represent the network weights.

\begin{enumerate}[leftmargin=0pt, itemindent=2pc, listparindent=\parindent]
	\item{\textit{Probability-Weighted Addition (PWA)}}:
	The predicted probability can be given by 
	\begin{equation}
		\label{Eq14}
		P_{P\!W\!A}=\alpha {{P}_{T\!F\!I}}+( 1-\alpha ){{P}_{Z\!C}},
	\end{equation}
	where \(P_{T\!F\!I}\) and \(P_{Z\!C}\) are the predicted probabilities obtained by \(F_{T\!F\!I}\) and \(F_{Z\!C}\) passing through fully connected (FC) layer and Softmax function. \(\alpha\) is the probability weight, which is set to 0.5.
	
	\item{\textit{Feature Vector Addition (FVA)}}:
	Element-wise addition is applied to \(F_{T\!F\!I}\) and \(F_{Z\!C}\), resulting in a new feature vector \(F_{F\!V\!A}\), which can be expressed as
	\begin{equation}
		\label{Eq15}
		F_{F\!V\!A}={{F}_{T\!F\!I}}+{{F}_{Z\!C}}.
	\end{equation}
	
	\(F_{F\!V\!A}\) retains the same size as the original feature vectors, and \(P_{F\!V\!A}\) can be calculated with \(F_{F\!V\!A}\).
	
	\item{\textit{Feature Vector Concatenation (FVC)}}:
	To preserve all the information from the original feature vectors, \(F_{T\!F\!I}\) and \(F_{Z\!C}\) are concatenated to form a new vector with a size twice that of the original ones, which can be given by 
	\begin{equation}
		\label{Eq16}
		P_{F\!V\!C}={{F}_{T\!F\!I}}\cup{{F}_{Z\!C}},
	\end{equation}
	where \(\cup\) denotes the concatenation operation. Then the predicted probabilities can be obtained by passing through FC layer and Softmax function.
\end{enumerate}

Let \(y=[y_1,y_2,\ldots,y_9]\) and \(\hat{y}=[\hat{y}_1,\hat{y}_2,\ldots,\hat{y}_9]\) denote the true and predicted labels, where \(y_i\) and \(\hat{y}_i\) equals 1 when the sample belongs to the \(i\text{-th}\) drone type; otherwise, they equal 0. Then the categorical cross-entropy loss function can be computed by
\begin{equation}
	\label{Eq17}
	\mathcal{L}=-\sum\limits_{i=1}^{9}{{{y}_{i}}}\log ({{\hat{y}}_{i}}),
\end{equation}
which can be utilized as the loss value for back-propagation and updating the network parameters.

\section{Numerical Results}
\label{sec4}
\subsection{Simulation Setup}

In DroneRFa, RF signals are randomly sampled based on the selected drone types as the raw data. Additive white Gaussian noise is then introduced over the background interference, with SNR varying from -15 dB to 15 dB in 2 dB intervals. This process resulted in a total of 23040 I/Q samples across 9 categories of RF signals, which are subsequently split into training, validation, and test sets in a 9:1:1 ratio. The hardware environment consists of an NVIDIA GeForce RTX 4060 Ti GPU and an 11th Gen Intel(R) Core(TM) i7-11700K @ 3.60GHz, with the remaining simulation parameters are provided in Table \ref{tab:table2}.

\begin{table}[!t]
	\caption{Simulation Parameters \label{tab:table2}}
	\centering
	\begin{tabular}{c c}
		\hline
		Parameter & Value \\
		\hline
		Batchsize for Sequence Feature-Based & 512 \\
		Batchsize for TFI Feature-Based & 256 \\
		Batchsize for Fusion Feature-Based & 128 \\
		Epoch for Sequence Feature-Based & 300 \\
		Epoch for TFI Feature-Based & 30 \\
		Epoch for Fusion Feature-Based & 30 \\
		Learning Rate & 0.0001 \\
		\hline
	\end{tabular}
\end{table}

The baseline and the proposed algorithms are given as follows.
\begin{enumerate}[leftmargin=0pt, itemindent=2pc, listparindent=\parindent]
	\item{\textit{I/Q Features-Based CNN (IQ-CNN)}}: I/Q sequences are adopted as input features for training the CNN.
	\item{\textit{NCPCS Features-Based CNN (NCPCS-CNN)}}: The NCPCS results computed from the I/Q sequences are used as input features for training the CNN.
	\item{\textit{Cross-Correlation Features for ZC Sequences-Based CNN (ZC-CNN)}}: The cross-correlation results between  I/Q samples and the locally generated ZC sequences are subjected to reduction and then utilized as input features for CNN's training.
	\item{\textit{TFI Features-Based MobileNetV3 (TFI-MobileNet)}}: The TFIs are computed and utilized for MobileNetV3's training.
	\item{\textit{PWV-Based Fusion Network (Fusion-PWV)}}: TFIs and the reduced cross-correlation results are processed through MobileNetV3 and CNN, respectively, to calculate the predicted probabilities. These probabilities are then weighted and fused to obtain the final predicted probabilities.
	\item{\textit{FVA-Based Fusion Network (Fusion-FVA)}}: MobileNetV3 and CNN output only feature vectors, which are combined through element-wise addition to generate a new vector with the same size, subsequently adopted for computing the prediction probabilities.
	\item{\textit{FVC-Based Fusion Network (Fusion-FVC)}}: For the feature vectors outputted by MobileNetV3 and CNN, they are concatenated to form a new vector that is twice the size of the original ones. This new vector is then utilized for calculating the prediction probabilities.
	
\end{enumerate}

\subsection{Simulation Results}

\begin{figure}[!t]
	\centering
	\includegraphics[width=2.5in]{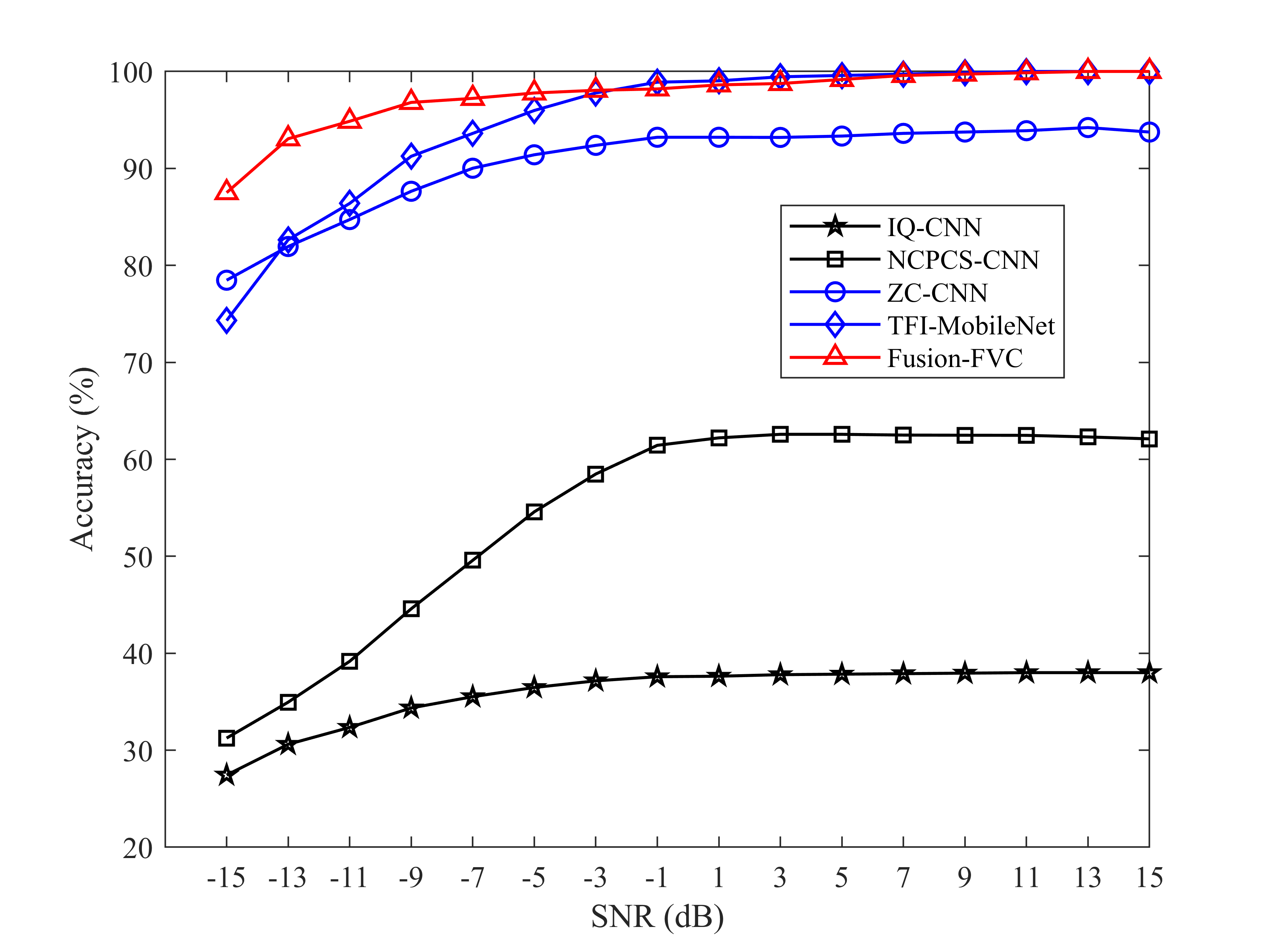}
	\caption{Drone RID accuracy of different algorithms.}
	\label{Fig_12}
\end{figure}
Fig. \ref{Fig_12} illustrates the drone RID accuracy of different algorithms. It can be observed that the accuracy increases with SNR, reaching a maximum of 100\% for the proposed algorithm. IQ-CNN exhibits the lowest performance, primarily because I/Q sequences struggle to directly reveal distinct and discriminative information under complex interference environments. NCPCS-CNN performs slightly better, owing to the inclusion of modulation parameters and frame structure information within the auto-correlation features. However, when \(f_s\) significantly exceeds the signal bandwidth, it becomes challenging to process a large volume of samples in real time to enhance NCPCS features. Furthermore, background noise also introduces interference and degradation to the auto-correlation peaks' amplitude.

On the other hand, although the features extracted by ZC-CNN are also susceptible to burst interference, the stronger correlation and robustness of ZC sequences ensure that the cross-correlation peaks retain most of their information. While TFI-MobileNet underperforms compared to ZC-CNN at low SNR, it achieves 100\% accuracy under favorable channel conditions. This is because TFI captures more comprehensive information, such as modulation parameters and frame structures of OFDM and FHSS signals. Additionally, TFI has lower computational cost, allowing for the inclusion of longer RF sampled signals, which ensures that RID remains accurate even when burst interference partially obscures the signal. The fusion of features based on ZC sequences and TFI demonstrates strong robustness under low SNR and burst interference conditions, while achieving RID accuracy of 100\% under high SNR conditions.

\begin{figure}[!t]
	\centering
	\includegraphics[width=2.5in]{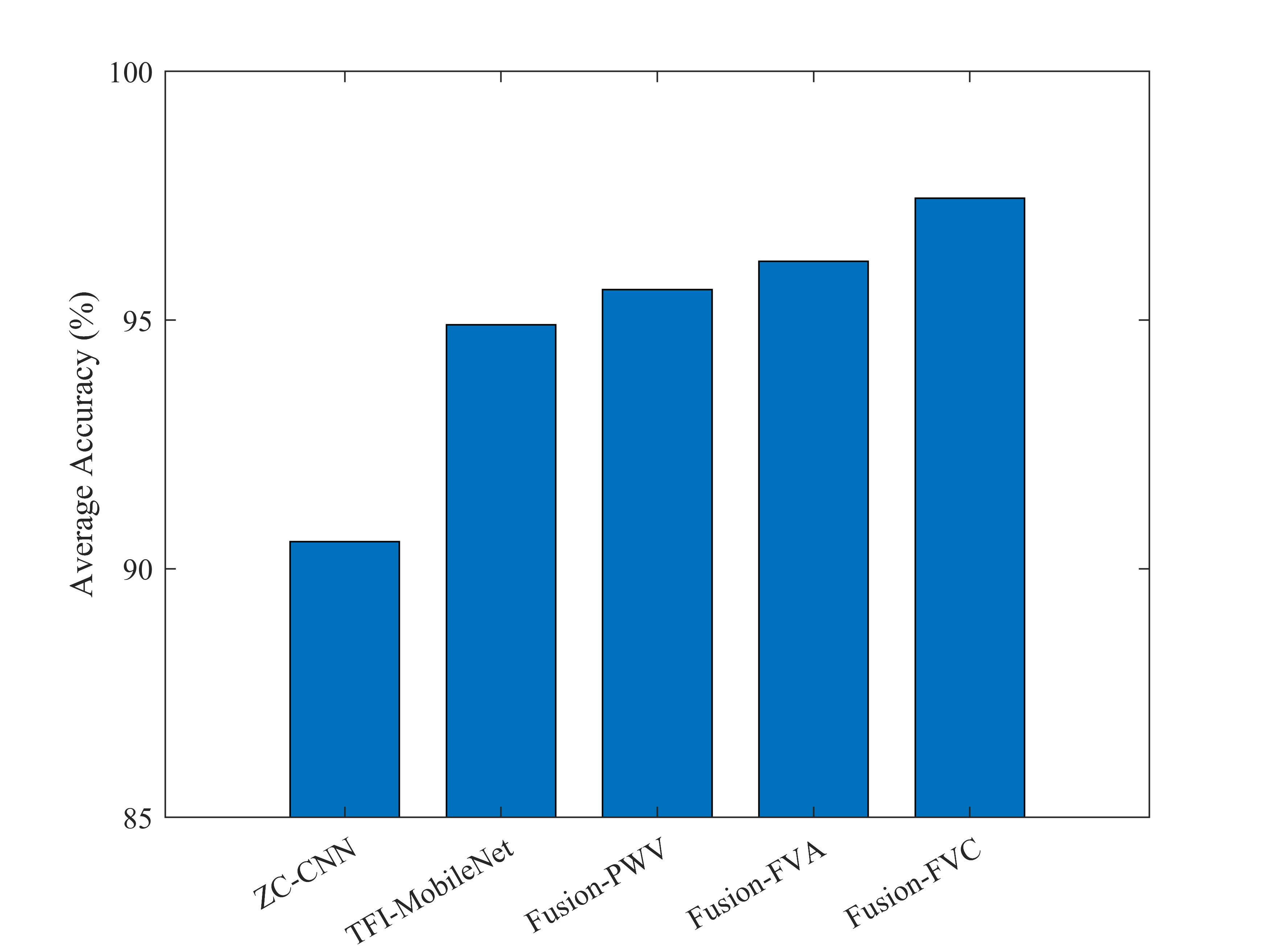}
	\caption{The impact of different feature fusion methods on RID accuracy.}
	\label{Fig_13}
\end{figure}

The impact of different feature fusion methods on the average accuracy is investigated in Fig. \ref{Fig_13}. Regardless of the fusion strategy employed, fused features consistently outperform single feature in terms of accuracy enhancement. Fusion-FVC achieves the best performance compared to PWV and FVA, as it retains the complete differential information embedded in the feature vectors, which improves the average RID accuracy by at least 2.5\%.

In addition, we present a comparison of the computational costs associated with different algorithms. Table \ref{tab:table3} indicates that feature fusion significantly increases computational costs, with PWV and FVA incurring higher overheads. This is primarily because feature concatenation relies on matrix operations, which are more efficient compared to branching computations and memory access.
\begin{figure}[!t]
	\centering
	\includegraphics[width=2.5in]{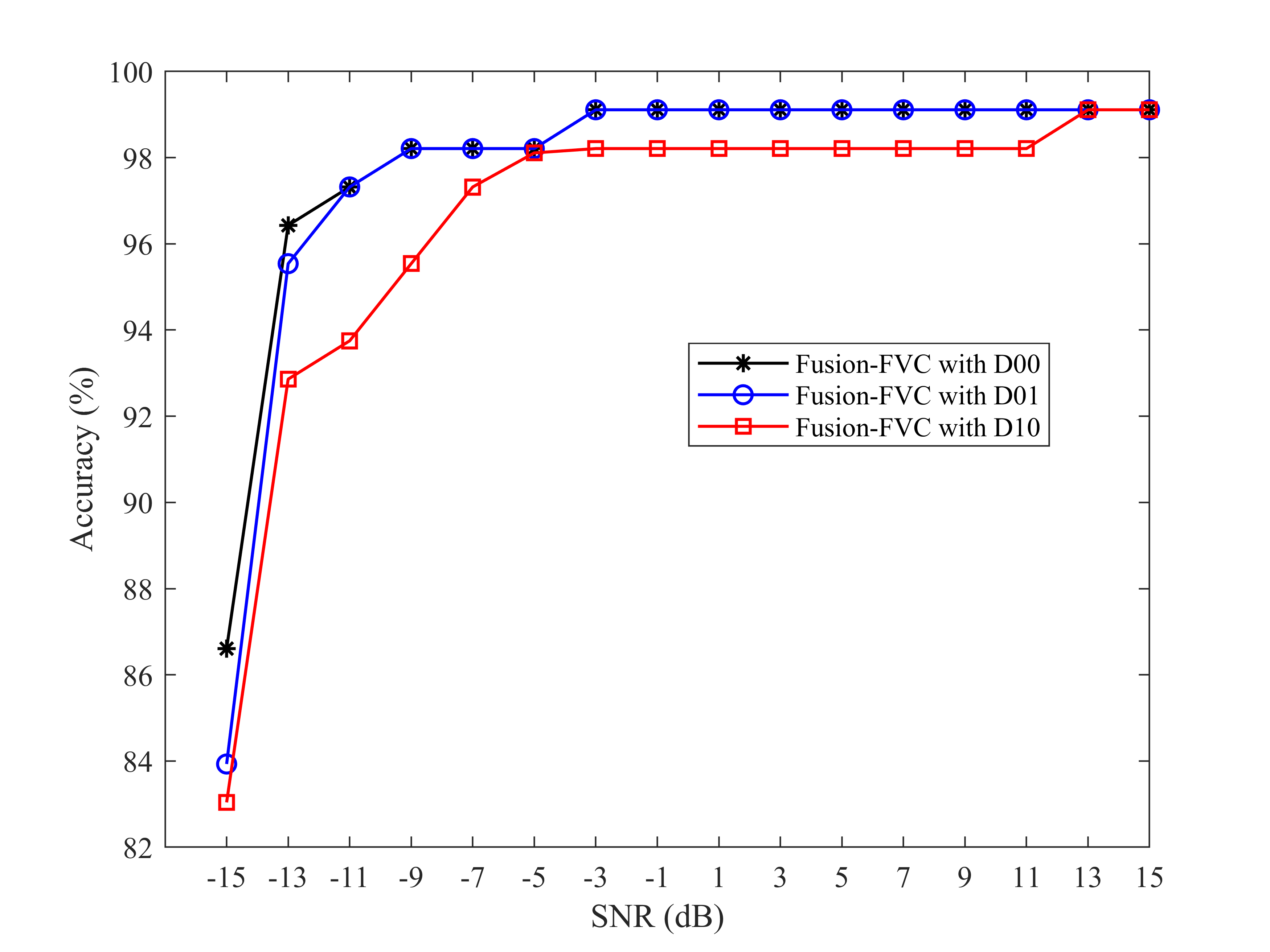}
	\caption{Drone RID accuracy versus flight distances.}
	\label{Fig_14}
\end{figure}
\begin{table}[!t]
	\caption{Computational Costs for Different Algorithms \label{tab:table3}}
	\centering
	\begin{tabular}{c c c}
		\hline
		Method & Average RID Accuracy & Computational Cost \\
		\hline
		ZC & 90.55\% & 2.47 GPU/h \\
		TFI & 94.91\% & 1.23 GPU/h \\
		Fusion-PWV & 95.62\% & 4.55 GPU/h \\
		Fusion-FVA & 96.18\% & 4.60 GPU/h \\
		Fusion-FVC & \textbf{97.28}\% & \textbf{3.70 GPU/h} \\
		\hline
	\end{tabular}
\end{table}

As only certain types of drones in DroneRFa include RF signals at varying flight distances, we analyzed the RID accuracy for a subset of drone types under different flight distances, as shown in Fig. \ref{Fig_14}. While the accuracy remains the same at high SNR, closer flight distances result in higher identification accuracy at low SNR. This is because longer distances lead to reduced energy in the received RF signals, which diminishes the distinguishing characteristics of both TFI and ZC sequence features.

\begin{figure}[!t]
	\centering
	\includegraphics[width=2.5in]{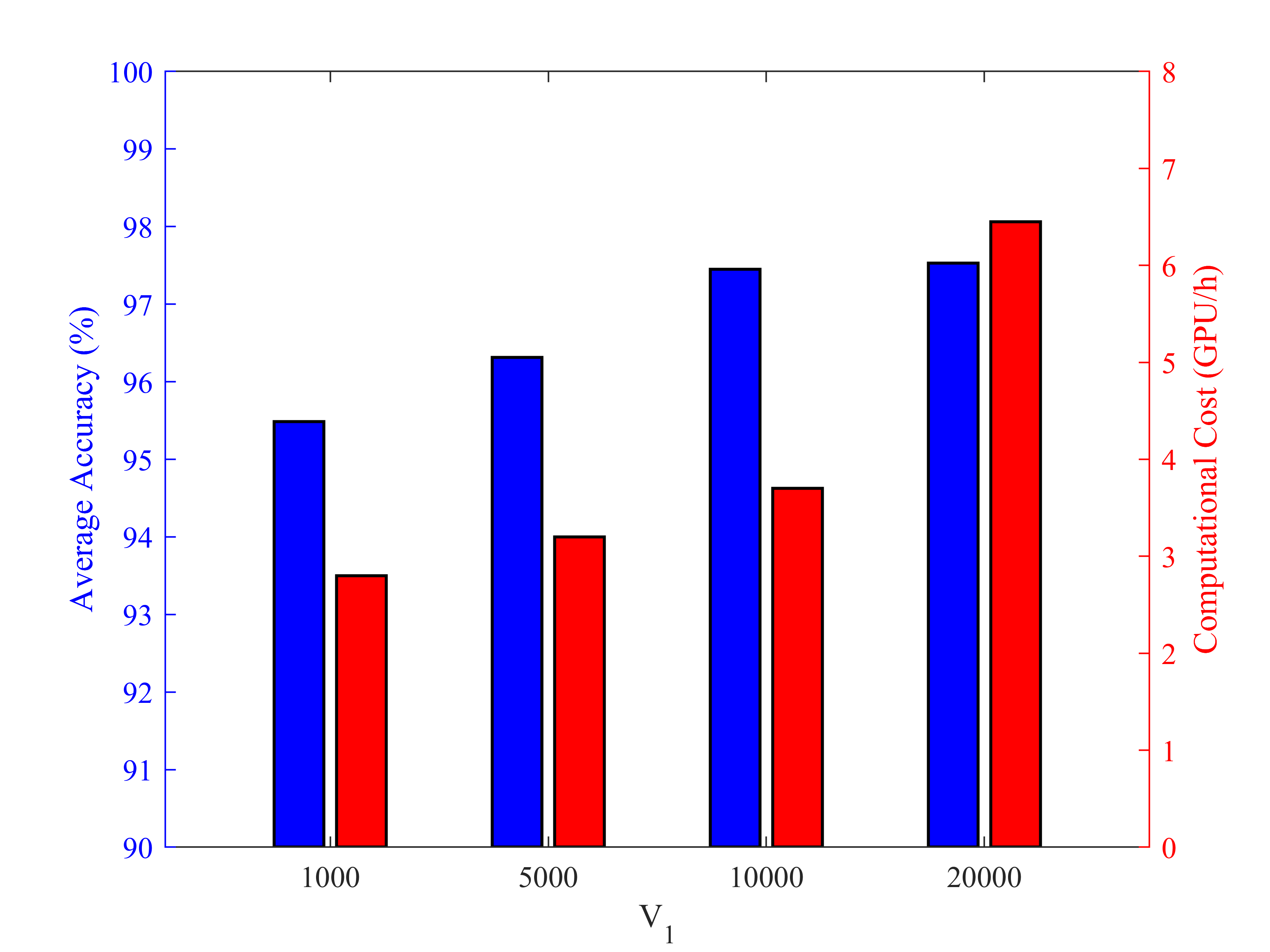}
	\caption{The average accuracy of Fusion-FVC versus \(\tiny{V_1}\).}
	\label{Fig_15}
\end{figure}
\begin{figure}[!t]
	\centering
	\includegraphics[width=2.5in]{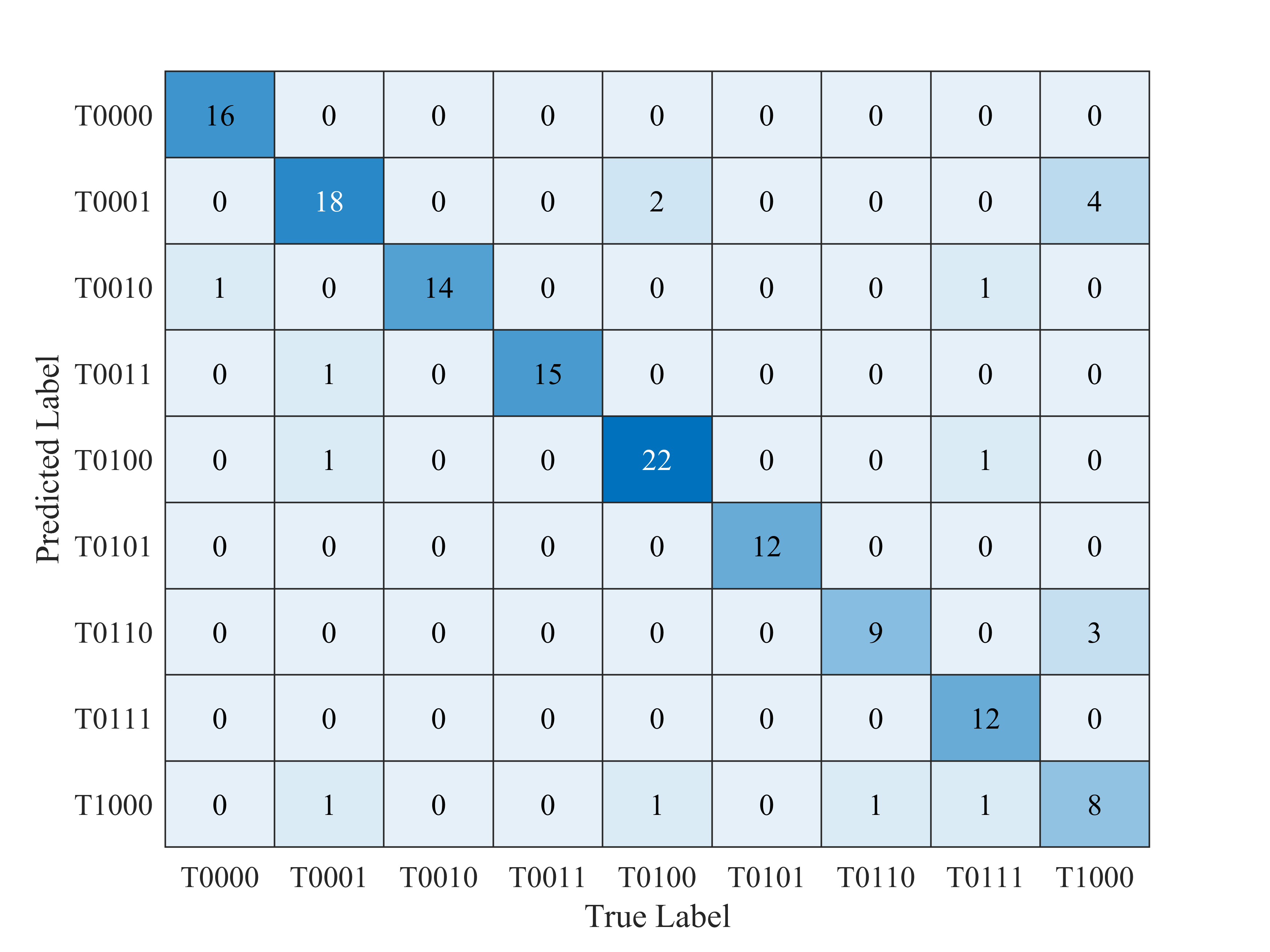}
	\caption{Confusion Matrix of Fusion-FVC with SNR = -15 dB.}
	\label{Fig_16}
\end{figure}
\begin{figure*}[!b]
	\centering
	\subfloat[]{\includegraphics[width=2.3in]{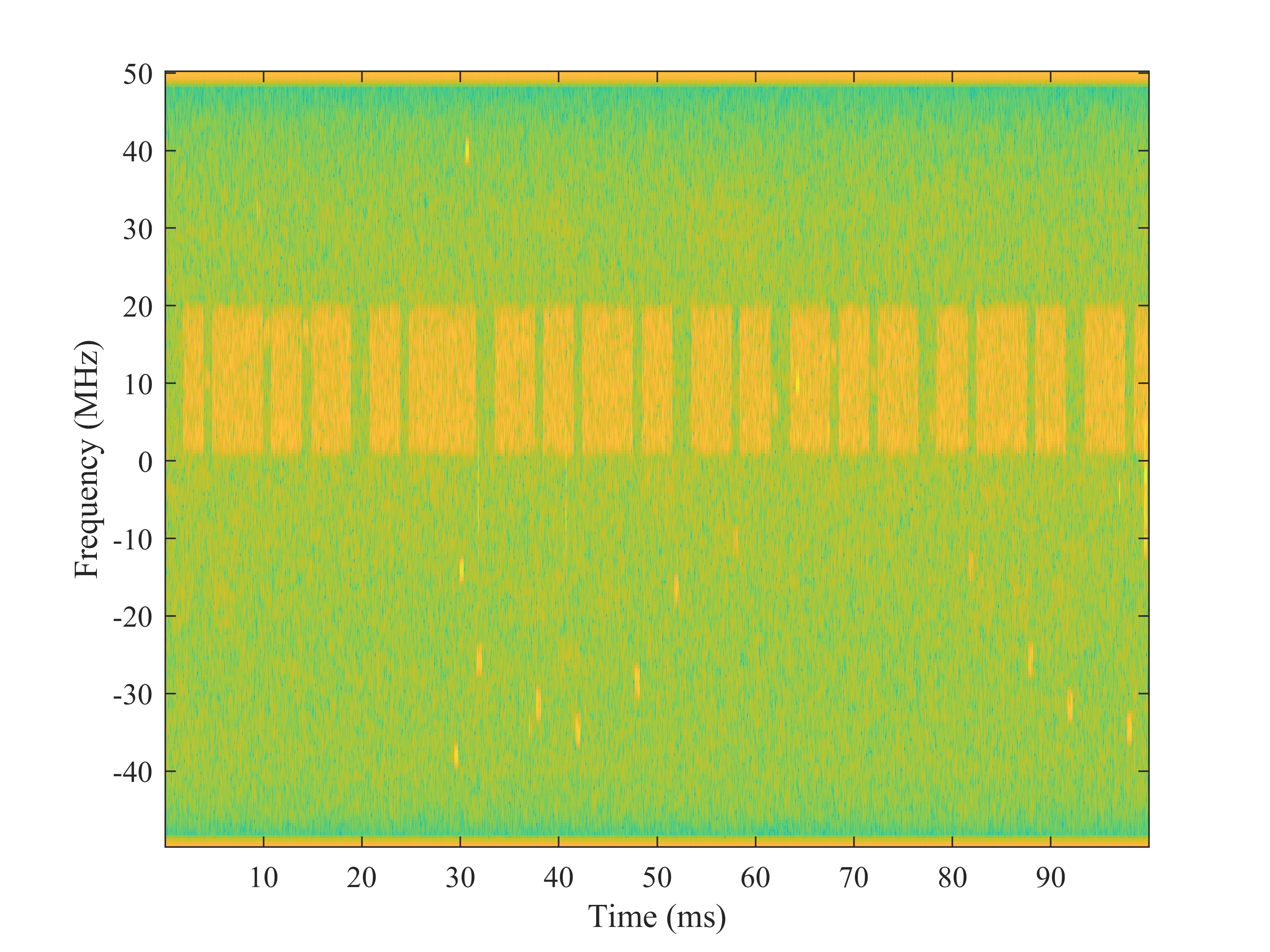}%
		\label{fig_first_case}}
	\hfil
	\subfloat[]{\includegraphics[width=2.3in]{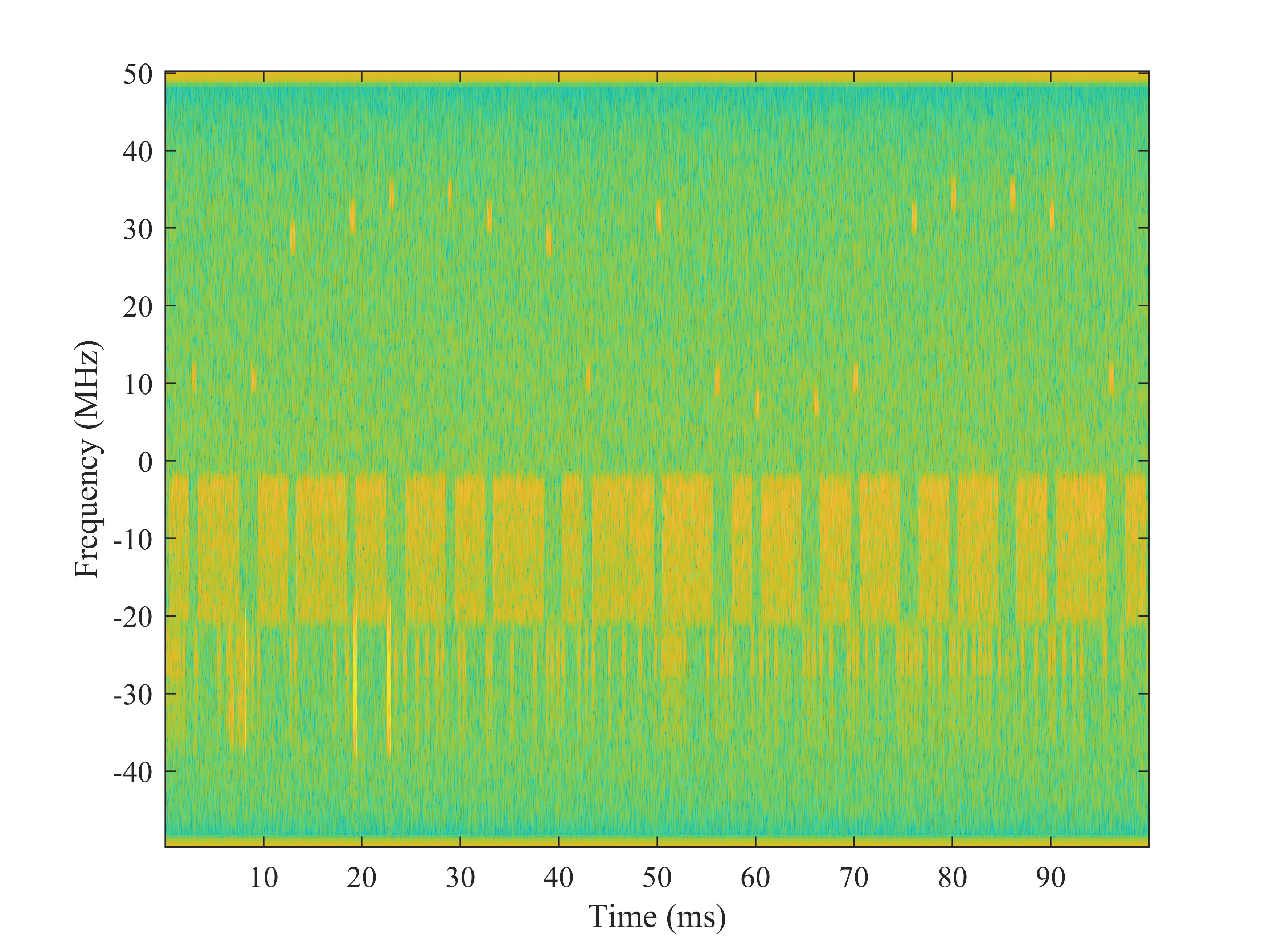}%
		\label{fig_second_case}}
	\hfil
	\subfloat[]{\includegraphics[width=2.3in]{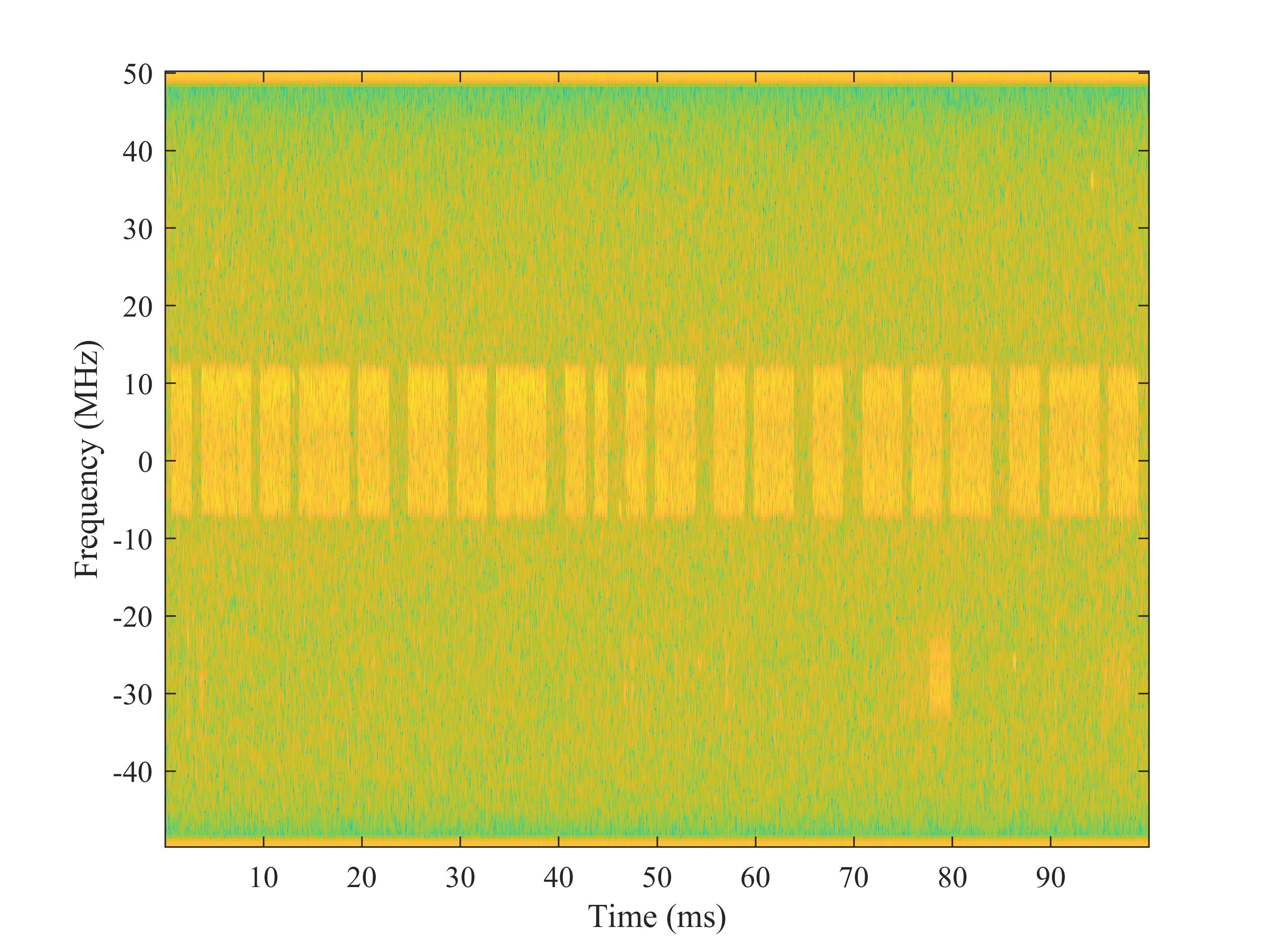}%
		\label{fig_third_case}}
	\caption{TFI of T000, T011, and T1000 with \(\tiny{10^7}\) samples at SNR = 15 dB. (a) T0001D00. (b) T0110D00. (c) T1000D00.}
	\label{Fig_17}
\end{figure*}

The value of \(V_1\) directly impacts the size of the ZC features, which subsequently influences the average accuracy of the proposed algorithm. Fig. \ref{Fig_15} illustrates that as \(V_1\) increases, the average accuracy of Fusion-FVC initially improves and then saturates, while the computational cost rises at an accelerating rate. When \(V_1=10^4\), the proposed algorithm achieves near-optimal average accuracy while avoiding the exponential growth in computational cost observed for \(V_1>10^4\). This validates the rationality of selecting \(V_1=10^4\) based on the frame structure and modulation parameters.

The confusion matrix of Fusion-FVC with SNR = -15 dB is shown in Fig. \ref{Fig_16}. It can be observed that T0001, T0110, and T1000 exhibit a certain degree of misclassification. This is primarily because the frame structures of these three types of drones present similar characteristics in TFI, as shown in Fig. \ref{Fig_17}, and the cross-correlation properties of ZC sequences become less distinguishable under extremely low SNR conditions.

\section{Conclusion}
\label{sec5}
This paper proposed a drone RID algorithm based on ZC sequences and TFI, enabling accurate identification of 8 types of drones with features extracted from RF signals in real-world scenarios. By analyzing the communication protocols of RF signals in the DroneRFa dataset, including modulation parameters and frame structures, prior knowledge of ZC sequences was obtained. The locally generated ZC sequences were cross-correlated with RF signals to extract ZC-based features. Considering that TFI is less affected by burst interference and encapsulates communication protocol information, TFI features were constructed from the RF signals. To reduce computational costs, data reduction was performed by analyzing the frame structures and parameters, minimizing the impact on feature performance. Three feature fusion methods, i.e., probability-weighted addition, feature vector addition, and feature vector concatenation, were employed, and their computational cost and accuracy were evaluated. Simulation results demonstrated that the proposed algorithm outperformed existing algorithms, achieving a minimum improvement of 2.5\% in RID accuracy. The proposed algorithm also maintained robust performance under low SNR, burst interference, and varying flight distances.


\begin{thebibliography}{99}
	\bibliographystyle{IEEEtran}
	
	\bibitem{ref1}
	V. Hassija et al., “Fast, reliable, and secure drone communication: A comprehensive survey,” {\it{IEEE Commun. Surveys Tuts.}}, vol. 23, no. 4, pp. 2802-2832, 4th Quart., 2021.
	
	\bibitem{ref2}
	Fortune Business Insights, “Unmanned aerial vehicle (UAV) market size,” https://www.fortunebusinessinsights.com/industry-reports/unmanned-aerial-vehicle-uav-market-101603, 2022, (Accessed: 2024-Nov-11).
	
	\bibitem{ref3}
	H. Zhang, T. Li, Y. Li, J. Li, O. A. Dobre, and Z. Wen, “RF-based drone classification under complex electromagnetic environments using deep learning,” {\it{IEEE Sensors J.}}, vol. 23, no. 6, pp. 6099-6108, 15 Mar. 2023.
	
	\bibitem{ref4}
	Y. Chen, L. Zhu, Y. Jiao, C. Yao, K. Cheng, and Y. Gu, “An extreme value theory-based approach for reliable drone RF signal identification,” {\it{IEEE Trans. Cogn. Commun. Netw.}}, vol. 10, no. 2, pp. 454-469, Apr. 2024.
	
	\bibitem{ref5}
	N. Yu, J. Wu, C. Zhou, Z. Shi, and J. Chen, “Open set learning for RF-based drone recognition via signal semantics,” {\it{IEEE Trans. Inf. Forensics Security}}, vol. 19, pp. 9894-9909, 2024.
	
	\bibitem{ref6}
	P. Shukla, S. Shukla, and A. K. Singh, “Trajectory-prediction techniques for unmanned aerial vehicles (UAVs): A comprehensive survey,” {\it{IEEE Commun. Surveys Tuts., early access}}, Oct. 1, 2024, doi: 10.1109/COMST.2024.3471671.
	
	\bibitem{ref7}
	L. Lv et al., “Safeguarding next-generation multiple access using physical layer security techniques: A tutorial,” {\it{Proc. IEEE}}, vol. 112, no. 9, pp. 1421-1466, Sep. 2024.
	
	\bibitem{ref8}
	K. Belwafi, R. Alkadi, S. A. Alameri, H. A. Hamadi, and A. Shoufan, “Unmanned aerial vehicles’ remote identification: A tutorial and survey,” {\it{IEEE Access}}, vol. 10, pp. 87577-87601, 2022.
	
	\bibitem{ref9}
	B. K. Kim, H.-S. Kang, and S.-O. Park, “Drone classification using convolutional neural networks with merged Doppler images,” {\it{IEEE Geosci. Remote Sens. Lett.}}, vol. 14, no. 1, pp. 38-42, Jan. 2017.
	
	\bibitem{ref10}
	B.-S. Oh, X. Guo, F. Wan, K.-A. Toh, and Z. Lin, “Micro-Doppler mini-UAV classification using empirical-mode decomposition features,” {\it{IEEE Geosci. Remote Sens. Lett.}}, vol. 15, no. 2, pp. 227-231, Feb. 2018.
	
	\bibitem{ref11}
	G. Fasano, D. Accado, A. Moccia, and D. Moroney, “Sense and avoid for unmanned aircraft systems,” {\it{IEEE Aerosp. Electron. Syst. Mag.}}, vol. 31, no. 11, pp. 82-110, Nov. 2016.
	
	\bibitem{ref12}
	J. Wang, Y. Liu, and H. Song, “Counter-unmanned aircraft system(s) (C-UAS): State of the art, challenges, and future trends,” {\it{IEEE Aerosp. Electron. Syst. Mag.}}, vol. 36, no. 3, pp. 4-29, Mar. 2021.
	
	\bibitem{ref13}
	X. Yue, Y. Liu, J. Wang, H. Song, and H. Cao, “Software defined radio and wireless acoustic networking for amateur drone surveillance,” {\it{IEEE Commun. Mag.}}, vol. 56, no. 4, pp. 90-97, Apri. 2018.
	
	\bibitem{ref14}
	Z. Shi, X. Chang, C. Yang, Z. Wu, and J. Wu, “An acoustic-based surveillance system for amateur drones detection and localization,” {\it{IEEE Trans. Veh. Technol.}}, vol. 69, no. 3, pp. 2731-2739, Mar. 2020.
	
	\bibitem{ref15}
	G. Reus-Muns and K. R. Chowdhury, “Classifying UAVs with proprietary waveforms via preamble feature extraction and federated learning,” {\it{IEEE Trans. Veh. Technol.}}, vol. 70, no. 7, pp. 6279-6290, Jul. 2021.
	
	\bibitem{ref16}
	H. Zhang, T. Li, N. Su, D. Wei, Y. Li, and Z. Wen, “Drone identification based on normalized cyclic prefix correlation spectrum,” {\it{IEEE Trans. Cogn. Commun. Netw.}}, vol. 10, no. 4, pp. 1241-1252, Aug. 2024.
	
	\bibitem{ref17}
	A. Alipour-Fanid, M. Dabaghchian, N. Wang, P. Wang, L. Zhao, and K. Zeng, “Machine learning-based delay-aware UAV detection and operation mode identification over encrypted Wi-Fi traffic,” {\it{IEEE Trans. Inf. Forensics Security}}, vol. 15, pp. 2346-2360, 2020.
	
	\bibitem{ref18}
	H. Kong, C. Huang, J. Yu, and X. Shen, “A survey of mmWave radar-based sensing in autonomous vehicles, smart homes and industry,” {\it{IEEE Commun. Surveys Tuts., early access}}, Jun. 11, 2024, doi: 10.1109/COMST.2024.3409556.
	
	\bibitem{ref19}
	M. H. Rahman, M. A. S. Sejan, M. A. Aziz, R. Tabassum, J.-I. Baik, and H.-K. Song, “A comprehensive survey of unmanned aerial vehicles detection and classification using machine learning approach: Challenges, solutions, and future directions,” {\it{Remote Sens.}}, vol. 16, no. 5, p. 879, 2024.
	
	\bibitem{ref20}
	Y.-C. Lai and T.-Y.Lin, “Vision-based mid-air object detection and avoidance approach for small unmanned aerial vehicles with deep learning and risk assessment,” {\it{Remote Sens.}},vol. 16, no. 5, p. 756, 2024.
	
	\bibitem{ref21}
	X. Zhou, G. Yang, Y. Chen, L. Li, and B. M. Chen, “VDTNet: A high-performance visual network for detecting and tracking of intruding drones,” {\it{IEEE Trans. Intell. Transp. Syst.}}, vol.25, no.8, pp.9828-9839, Aug. 2024.
	
	\bibitem{ref22}
	J. Chen et al., “Low-altitude UAV surveillance system via highly sensitive distributed acoustic sensing,” {\it{IEEE Sensors J.}}, vol. 24, no. 20, pp. 32237-32246, Oct. 2024.
	
	\bibitem{ref23}
	C. J. Swinney and J. C. Woods, “A review of security incidents and defence techniques relating to the malicious use of small unmanned aerial systems,” {\it{IEEE Aerosp. Electron. Syst. Mag.}}, vol. 37, no. 5, pp. 14-28, May 2022.
	
	\bibitem{ref24}
	M. A. Khan, H. Menouar, A. Eldeeb, A. Abu-Dayya, and F. D. Salim, “On the detection of unauthorized drones—techniques and future perspectives: A review,” {\it{IEEE Sensors J.}}, vol. 22, no. 12, pp. 11439-11455, Jun. 2022.
	
	\bibitem{ref25}
	H. Dong, J. Liu, C. Wang, H. Cao, C. Shen, and J. Tang, “Drone detection method based on the time-frequency complementary enhancement model,” {\it{IEEE Trans. Instrum. Meas.}}, vol.72, pp.1-12, Oct. 2023.
	
	\bibitem{ref26}
	M. Ezuma, F. Erden, C. Kumar Anjinappa, O. Ozdemir, and I. Guvenc, “Detection and classification of UAVs using RF fingerprints in the presence of Wi-Fi and bluetooth interference,” {\it{IEEE Open J. Commun. Soc.}}, vol. 1, pp. 60-76, 2020.
	
	\bibitem{ref27}
	A. Gumaei et al., “Deep learning and blockchain with edge computing for 5G-enabled drone identification and flight mode detection,” {\it{IEEE Netw.}}, vol. 35, no. 1, pp. 94-100, Jan. 2021.
	
	\bibitem{ref28}
	W. Nie et al., “UAV detection and localization based on multi-dimensional signal features,” {\it{IEEE Sensors J.}}, vol. 22, no. 6, pp. 5150-5162, Mar. 2022.
	
	\bibitem{ref29}
	R. Zhao, T. Li, Y. Li, Y. Ruan, and R. Zhang, “Anchor-free multi-UAV detection and classification using spectrogram,” {\it{IEEE Internet Things J.}}, vol. 11, no. 3, pp. 5259-5272, Feb. 2024.
	
	\bibitem{ref30}
	Z. Cai, Y. Wang, Q. Jiang, G. Gui, and J. Sha, “Toward intelligent lightweight and efficient UAV identification with RF fingerprinting,” {\it{IEEE Internet Things J.}}, vol. 11, no. 15, pp. 26329-26339, Aug. 2024.
	
	\bibitem{ref31}
	R. Akter, V.-S. Doan, A. Zainudin, and D.-S. Kim, “Sparsely connected low complexity CNN for unmanned vehicles detection-sensing RF signal,” {\it{IEEE Trans. Veh. Technol.}}, vol. 73, no. 10, pp. 14236-14251, Oct. 2024.
	
	\bibitem{ref32}
	S. Basak, S. Rajendran, S. Pollin, and B. Scheers, “Combined RF-based drone detection and classification,” {\it{IEEE Trans. Cogn. Commun. Netw.}}, vol. 8, no. 1, pp. 111-120, Mar. 2022.
	
	\bibitem{ref33}
	Z. Wang, Z. Cao, J. Xie, W. Zhang, and Z. He, “RF-based drone detection enhancement via a generalized denoising and interference-removal framework,” {\it{IEEE Singal Process. Lett.}}, vol.31, pp.929-933, Mar. 2024.
	
	\bibitem{ref34}
	Y. Chen, L. Zhu, J. Zhang, K. Cheng, L. Yu, and Y. Gu, “Counterfactual threshold learning for drone RF signal classification under interference conditions,” {\it{IEEE Wireless Commun. Lett.}}, vol. 13, no. 7, pp. 1958-1962, Jul. 2024.
	
	\bibitem{ref35}
	S. Basak, S. Rajendran, S. Pollin, and B. Scheers, “Spectrum Prediction for Protocol-Aware RF Jamming,” {\it{IEEE Trans. Cogn. Commun. Netw.}}, vol.10, no.2, pp.363-373, Oct. 2024.
	
	\bibitem{ref36}
	C. Bender, “DJI drone IDs are not encrypted,” 2022, {\it{arXiv:}} 2207.10795.
	
	\bibitem{ref37}
	I. Bisio, C. Garibotto, F. Lavagetto, A. Sciarrone, and S. Zappatore, “Blind detection: Advanced techniques for WiFi-based drone surveillance,” {\it{IEEE Trans. Veh. Technol.}}, vol. 68, no. 1, pp. 938-946, Jan. 2019.
	
	\bibitem{ref38}
	N. Schiller et al., “Drone Ssecurity and the mysterious case of DJI’s droneID,” {\it{Proc. Netw. Distrib. Syst. Security Symp. (NDSS)}}, 2023, pp. 1-17.
	
	\bibitem{ref39}
	N. Yu, S. Mao, C. Zhou, G. Sun, Z. Shi, and J. Chen, “DroneRFa: A large-scale dataset of drone radio frequency signals for detecting low-altitude drones,” {\it{J. Electron. Inf. Technol.}}, vol. 46, no. 4, pp. 1147-1156, Apr. 2024.
	
	\bibitem{ref40}
	D. G. Manolakis, V. K. Ingle, and S. M. Kogon, {\it{Statistical and Adaptive Signal Processing.}}. New York: McGraw-Hill, 2000.
	
\end{thebibliography}
\end{document}